\documentclass{article}
\usepackage[utf8]{inputenc}
\usepackage{authblk}
\usepackage{setspace}
\usepackage[margin=1.25in]{geometry}
\usepackage{graphicx}
\graphicspath{ {./figures/} }
\usepackage{subcaption}
\usepackage{multirow}
\usepackage{algorithm}
\usepackage{algpseudocode}
\usepackage{amsmath}
\usepackage{xcolor}
\usepackage{hyperref}

\title{A Gaussian-based spatially weighted loss formulation for Physics-Informed Neural Networks with Shocks}

\author[1*]{Kheir-eddine Otmani}
\author[2]{Abdelhalim Azzouz}
\author[3]{Nourelhouda Groun}
\author[4]{Esteban Ferrer}

\affil[1]{Department of Mathematics, University Center of El Bayadh Nour El Bachir, El Bayadh, Algeria}
\affil[2]{Department of Mathematics, University Salhi Ahmed, Naama, Algeria}
\affil[3]{Department of Mathematics, University Mohamed Khider Biskra, Biskra, Algeria.}
\affil[4]{ETSIAE-UPM - School of Aeronautics, Universidad Politécnica de Madrid, Plaza Cardenal Cisneros 3, E-28040, Madrid, Spain}
\affil[*]{Address correspondence to: k.otmani@cu-elbayadh.dz}
\date{}

\begin{document}

\maketitle
\begin{abstract}
Physics-informed neural networks (PINNs) often struggle to resolve shocks because smooth neural representations and globally averaged residual losses tend to underemphasize localized high-gradient regions. We introduce a residual-tracked Gaussian weighting strategy that represents the spatial distribution of the PDE-residual loss through a small number of learnable parameters, without resampling the collocation points or prescribing the shock location. The Gaussian center and width are updated from the evolving PDE residual, allowing the weighting field to follow stationary and moving shocks during training without prior knowledge of their positions or trajectories. The method is assessed using one-dimensional low-viscosity Burgers problems and a two-dimensional Euler Riemann problem with four interacting shocks. For $\nu=5\times10^{-4}$, the relative $L_2$ error decreases from approximately $45\%$ to $13\%$ for the stationary shock and from $33\%$ to $14\%$ for the moving shock, compared with a standard PINN. In the two-dimensional problem, the maximum absolute density error decreases from $1.08$ to $0.62$. These results show that a compact, dynamically localized residual weighting can substantially improve PINN resolution of shock-dominated solutions while retaining the original collocation set and avoiding prescribed shock trajectories.
\end{abstract}


\tableofcontents

\section{Introduction}
Partial differential equations (PDEs) serve as fundamental tools for describing complex systems in a wide range of disciplines, including cellular dynamics in biology, macroscopic traffic congestion, fluid mechanics, and stochastic fluctuations in financial markets. In many nonlinear PDEs, the solution may develop steep gradients, sharp discontinuities, or shock-like structures, even from smooth initial conditions; examples include shock waves in compressible flows, congestion fronts in traffic-flow models, moving interfaces in biological transport, and abrupt transitions in financial or stochastic conservation-law models.

While traditional numerical schemes (e.g., finite differences, finite elements) have been for a long time the standard choice to solve PDEs, recently deep learning models and specifically Physics Informed Neural Networks (PINNs) are gaining traction due to their ability to integrate physical laws derived from the PDEs residuals directly all within a mesh-free environment. Unlike purely data-driven models, which act as a black box by interpolating between data points, PINNs provide physically consistent solutions with the governing equations. Despite their advantages, PINNs often failed to accurately resolve discontinuities and shock wave regimes. Because PINNs are fundamentally built using smooth activation functions, they exhibit a spectral bias toward low-frequency and smooth solutions. Consequently, they fail to capture sharp gradients and discontinuities. 

To address these difficulties, recent research has increasingly focused on adaptive refinement strategies that redistribute collocation points to regions characterized by sharp gradients, discontinuities, and shock waves. These limitations have motivated a broad class of refinement techniques aimed at improving both the approximation capability and the training efficiency of PINNs. A first line of work focuses on data and sampling refinement, including residual-based adaptive refinement (RAR), which focuses on  intelligently adding more collocation points in regions where the PDE solution is difficult to learn. This approach was used in several researches such as that of Liu et al.   \cite{liu2024adaptive}, where they introduced a novel adaptive sampling algorithm EI-RAR that increases the focus on sample points at the boundaries of the solution domain. The authors select a residual neural network and combine it with adaptive sampling algorithms for a series of numerical experiments, with numerical results indicating that, with the same number of residual points, the EI-RAR algorithm is more precise compared to other sampling methods. Tian et al. \cite{tian2025residual} proposes a Residual-Based Adaptive Refinement Physics-Informed Neural Networks (RAR-PINNs) method, which synergizes the nonlinear approximation capability of PINNs with a residual-driven adaptive sampling strategy. The authors demonstrate the efficiency of their tool using numerical experiments on two variants of the fifth-order KdV equation demonstrated, where  RAR-PINNs significantly outperform conventional PINNs in terms of both accuracy and computational efficiency. Similar approaches employing residual-based adaptive refinement are introduced in \cite{song2024physics}, which introduces G-RAS and its application on six widely used benchmarks, demonstrating that G-RAS can improve prediction accuracy and convergence speed compared to 10 SOTA methods, and also \cite{hanna2022residual} where a novel residual-based adaptive PINN is developed for a two-phase flow problem in porous media. Another widely used approach is  adaptive collocation strategies that concentrate training points in regions of high error or complex dynamics. Visser et al. \cite{visser2026pacmann} present a Point Adaptive Collocation Method for Artificial Neural Networks (PACMANN), which moves collocation points toward regions with higher residuals using gradient-based optimization algorithms guided by the gradient of the PINNs loss function. The authors apply their approach to several forward and inverse problems, including a low-regularity solution case and the 3D Navier–Stokes equations. The results show that the method achieves state-of-the-art accuracy–efficiency performance for low-dimensional problems and exceeds existing approaches in high-dimensional settings. Celaya et al. \cite{celaya2025adaptive} proposed two adaptive collocation point selection strategies utilizing the QR Discrete Empirical Interpolation Method (QR-DEIM), a reduced-order modeling technique to efficiently approximate nonlinear functions. Their approach has proven to be successful in improving the accuracy of PINNs compared to existing methods, offering a promising direction for adaptive collocation point strategies. Other researchers have employed approaches such as that of Li et al. \cite{li2025adapw} and Lu et al. \cite{lu2026r} Effectively reduces large ‐ error points early and improves training performance and achieves faster convergence with fewer points, which is superior for problems with localized large errors in their researchers, respectively.

In parallel, domain decomposition methods, such as extended PINNs (XPINNs) proposed by Jagtap et al. \cite{jagtap2020extended}  and hp-adaptive variants introduced by Kharazmi et al. \cite{kharazmi2021hp}, divide the computational domain into subregions to allow localized learning, better scalability, and targeted refinement. Other approaches emphasize architectural enhancements, including adaptive activation functions \cite{jagtap2020adaptive,zhang2025simple}, Fourier feature embeddings \cite{wang2021eigenvector,jeong2025fourier}, and sinusoidal networks \cite{sitzmann2020implicit}, which aim to mitigate spectral bias and improve the representation of high-frequency solution components. Additionally, multi-fidelity and hybrid methods integrate coarse and fine data or combine PINNs with classical numerical solvers (e.g., finite element or finite difference methods) to enhance robustness and accuracy. For shock capturing problems, Han et al. \cite{HAN2025309} introduced an 'RH-piecewise PINN' framework that combines domain-decomposition methods with the Rankine-Hugoniot (RH) jump conditions to accurately capture sharp, non-linear shock waves without relying on dense, resource-heavy grid sampling. Concurrently, Li et al. \cite{Li2026_ARHPINN} introduced an adaptive regularized hybrid framework (ARH-PINN) combining Fourier embeddings and RBF layers to resolve high spatial gradients while suppressing non-physical numerical oscillations in hyperbolic conservation laws.

Recently, increasing attention has been directed toward the loss function itself as a central source of inefficiency. In standard PINNs, the loss is composed of multiple competing terms—typically enforcing PDE residuals, boundary conditions, and initial constraints—whose imbalance can lead to gradient stiffness and poor optimization trajectories. To address this issue, a variety of loss function refinement and adaptive weighting strategies have been proposed. These include heuristic and rule-based weighting. Researchers such as Braga-Neto et al. \cite{braga2021self} provided in their paper a definitive diagnostic that explains why weighting is necessary. More importantly, they introduce Learning Rate Annealing (LRA), which serves as a direct bridge between this theory and a practical adaptive weighting strategy. Meanwhile, Heydari et al. \cite{heydari2019softadapt} proposed a straightforward heuristic family of algorithms that dynamically adjusts the weights of different loss components based on their live performance statistics. on the other hand, Cao et al. \cite{cao2025wbpinn}  introduced an adaptive weighting algorithm that avoids the redundancy of manual parameter tuning by incorporating a correlation loss term and a penalty term. Their method significantly outperforms standard PINN and other recent methods. Gradient-based normalization techniques are also designed to tackle a key challenge in training PINNs, which is balancing the competing objectives encoded in their multi-part loss functions. Mochalin et al. \cite{mochalin2025enhancement} proposed a novel deep learning algorithm within the framework of the physics-informed neural network (PINN) architecture. The authors adapt and integrate a procedure, known as GradNorm, into the PINN framework to adjust the weight configuration to equalize the gradients of the components of the loss function. The application of the new algorithm shows great potential compared to conventional PINN. Lei et al. \cite{lei2026ddr} introduced a dynamic domain–gradient loss reweighting PINNs (DDR-PINN), which introduces a dual-residual reweighting mechanism based on gradient variations. The application of their technique not only shows powerful results but also consistently outperforms the standard PINN, APINN, and VI-PINNs with the fewest trainable parameters. Complementary ideas such as curriculum learning \cite{zhou2026physics} and constraint annealing \cite{abbas2025multi,son2023enhanced} progressively introduce or emphasize different loss terms during training. Compared to purely sampling-based refinement, these methods directly target the optimization landscape, offering a principled and often computationally efficient way to stabilize training and improve convergence.
Recent advances, such as: Liu et al. \cite{CGPINNS} propose a dynamic temporal weighting scheme that enforces causality by progressively shifting the training focus from early to late time steps. This ensures that the model respects the chronological causal structure of physical systems. 

Our work proposes a new alternative to adaptive resampling and domain decomposition methods, our work introduces a Gaussian-based spatially weighted loss formulation (G-PINNs), by modulating the PDE residual with a Gaussian distribution that tracks the discontinuity dynamically during the training process, the model selectively penalizes regions with higher probability of sharp gradients and shock waves. This allows the model to concentrate the optimization effort on resolving the evolution of the shock. In the following sections, we demonstrate the effectiveness of G-PINNs for both one- and two-dimensional problems.

The remainder of this paper is organized as follows. Section \ref{sec: meth} presents the proposed G-PINN formulation. Section \ref{sec: results} evaluates the method using three one-dimensional low-viscosity Burgers test cases, ranging from stationary to moving shock-wave regimes, together with a two-dimensional Euler Riemann problem involving multiple interacting shocks. Finally, Section \ref{sec: conclusion} summarises the main conclusions and provides directions for future work.

\section{Methodology} \label{sec: meth}
This section details G-PINNs: the Gaussian-based spatially weighted loss formulation. We describe the integration of the dynamic Gaussian weighting framework with the standard PINN solver to facilitate tracking of moving discontinuities.
\subsection{Physics Informed Neural Networks (PINNs)}\label{meth:pinns}
Consider the initial- boundary value problem:
\begin{equation} \label{eq: PDE}
\begin{cases}
 u_t + \mathcal{F}[u] = 0, \quad x\in \Omega , \quad t\in [0, T] \\
 u(x,0) = u_0, \quad x \in \Omega \\
 \mathcal{B}[u] = 0, \quad x \in \partial \Omega,
\end{cases}    
\end{equation}
where $\mathcal{F}$ is a differential operator (possibly non-linear), and $\mathcal{B}$ is the boundary conditions operator embedding Dirichlet, Neumann or mixed boundary conditions. To solve this type of problem, Raissi et al. \cite{RAISSIPINNS} introduced PINNs, a mesh-free deep learning framework in which an artificial neural network is trained to minimize the following loss objective:
\begin{equation} \label{eq: loss_standard}
\mathcal{L}(\theta) = \omega_{pde}\mathcal{L}_{pde}(\theta) + \omega_{ic} \mathcal{L}_{ic}(\theta) +\omega_{bc} \mathcal{L}_{bc}(\theta)    
\end{equation}

where:
\begin{itemize}
    \item $\mathcal{L}_{pde}(\theta) = \frac{1}{N_{int}} \sum_{i=0}^{N_{int}} R^2$, and $R= |\hat{u}_t^\theta(x^i,t^i) - \mathcal{F}[\hat{u}^\theta](x^i,t^i)|$ is the PDE residual,
    \item $\mathcal{L}_{ic}(\theta) = \frac{1}{N_{ic}} \sum_{i=0}^{N_{ic}} |\hat{u}^\theta(x^i,t^i) - u_0(x^i,t^i)|^2$ 
    \item $\mathcal{L}_{bc}(\theta)=\frac{1}{N_{bc}} \sum_{i=0}^{N_{bc}} |\hat{u}^\theta(x^i,t^i) - \mathcal{B}[\hat{u}^\theta](x^i,t^i)|^2$,
    \item $\omega_{pde}$, $\omega_{ic}$ and $\omega_{bc}$ are adaptive weights used to balance the loss term contributions. 
\end{itemize}
$\mathcal{L}_{pde}$ is the residual loss that quantifies the deviation of the predicted $\hat{u}^\theta$ from the governing PDE, $\mathcal{L}_{ic}$ and $\mathcal{L}_{bc}$ are the loss of the initial and boundary data that measure the discrepancy between the prediction of the neural network and the initial and boundary conditions, respectively.

\subsection{Gaussian-based spatially weighted loss formulation} \label{subsec: G-pinn formulation}
In this work, instead of using the standard loss formulation presented in Section~\ref{meth:pinns}, we propose the following residual loss formulation:
\begin{equation} \label{eq: GSPINN-Loss}
    \mathcal{L}_{pde}^\phi(\theta) = \frac{1}{N_{int}} \sum_{i=1}^{N_{int}} \phi(x^i,t^i) .R_i^2 
\end{equation}
where: $\phi$ is a normalized Gaussian probability density parametrized with the mean $\mu$ and standard deviation $\sigma$; and $\theta$ are the weights of the neural network.

For each collocation point $(x^i,t^i)$, an non-normalized Gaussian density is first defined as
\begin{equation}
\label{eq:gaussian_score}
g_i(\theta_G)
=
\frac{1}{\sigma\sqrt{2\pi}}
e^
-\frac{x^i-\mu^2}
{2\sigma^2}.
\end{equation}
with $\theta_G =\{\mu, \sigma\}$.

To enable dynamical spatial weighting, the mean and standard deviation are parameterized as
\begin{align}
\mu(t) &= mt+c,
\label{eq:mean}
\end{align}
\begin{align}
\sigma(t) &=
\ln(1+e^{wt+b}),
\label{eq:deviation}
\end{align}

The function $\sigma(t) =
\ln(1+e^{wt+b})$ is used to guarantee that the standard deviation remains strictly positive during training. In the numerical implementation, practical lower and upper bounds are imposed on the standard deviation to prevent excessively narrow or diffuse Gaussian distributions and to maintain numerical stability during optimization.

This formulation is beneficial for problems involving steep moving gradients, such as time propagating shock waves. If $\mu(t)$ corresponds to the center of the shock wave $x_s$ at time $t$, then the learned propagation speed of the moving shock wave is given by
$$
\frac{d \mu}{dt} = m,
$$
while $c$ denotes the initial position of the shock wave. Similarly, the evolution of $\sigma(t)$ through the parameters $w$ and $b$ allows the residual loss to adapt to the sharpening of the shock wave. By dynamically fitting the parameters $m$, $c$, $w$ and $b$ during training, the Gaussian distribution tracks the underlying PDE physics, ensuring that the collocation points in high-gradient regions are consistently prioritized during optimization.

To prevent the Gaussian amplitude from arbitrarily changing the magnitude of the residual loss, the Gaussian density is normalized over the collocation points according to
\begin{equation}
\label{eq:normalized_gaussian}
\phi_i(\theta_G)
=
\frac{g_i(\theta_G)}
{\frac{1}{N_{int}}\sum_{j=1}^{N_{int}}g_j(\theta_G)}.
\end{equation}
Consequently,
\begin{equation}
\frac{1}{N_{int}}\sum_{i=1}^{N_{int}}\phi_i=1,
\end{equation}
so that the Gaussian weighting changes the spatial distribution of the loss contribution without introducing an arbitrary global scaling of the loss.

The new complete loss of the network becomes:
\begin{equation} \label{eq: L_modified}
    \mathcal{L}(\theta_G,\theta) = \omega_{pde}\mathcal{L}^{\phi}_{pde}(\theta_G, \theta) + \omega_{ic} \mathcal{L}_{ic}(\theta) +\omega_{bc} \mathcal{L}_{bc}(\theta).
\end{equation}

\subsubsection{Residual-based Gaussian Tracking}

The Gaussian distribution $\phi$ is parametrized through the functions defined in equations \ref{eq:mean} and \ref{eq:deviation} designed to track the maximum residual loss, this allows the Gaussian distribution to be dynamically centered in regions of steep gradients. Parameters $\theta_G = \{ m,c,b,w\}$ are optimized to minimize the following objective loss:
\begin{equation} \label{eq: loss_g}
   \mathcal{L}_G(\theta_G) = - \frac{1}{N_{int}} \sum_{i=1}^{N_{int}} \phi(x^i,t^i,
\theta_G). R^2,
\end{equation}

where the parameters $\theta_G$ are updated across training iterations. 
The output $\phi$ of this linear regression is then used in the loss formulation \ref{eq: GSPINN-Loss}. Notice that $\mathcal{L}_{G}(\theta_G) = -\mathcal{L}^{\phi}_{PDE}$, this means that while the network weights are optimized to minimize the residual $\mathcal{L}^{\phi}_{PDE}$, the Gaussian parameters $\theta_G$ are updated to maximize this same residual. Consequently, $\mathcal{L}_{G}$ automatically aligns the Gaussian distributions with shock fronts and discontinuities where the residual is highest.


\subsection{Training strategy}

The training strategy consists of an initial standard PINN training phase followed by an alternating optimization phase.

\begin{enumerate}
\item \textbf{Initialization:} Training is first performed for $N_{init}$ epochs using the standard PINN loss. This initial phase allows the neural network to develop a meaningful approximation of the PDE solution before the residual-based Gaussian tracking mechanism is activated.

\item \textbf{Synchronous training:} After initialization, the neural network parameters and Gaussian parameters are optimized alternately. At each training iteration, the following two steps are performed:

\begin{itemize}
    \item \textbf{Residual-based Gaussian tracking:} The neural network parameters $\theta$ are temporarily fixed, and the Gaussian parameters $\theta_G=\{m,c,w,b\}$ are updated to minimize $\mathcal{L}_G$. Since the residual is fixed during this step, the Gaussian distribution adapts its center and width to emphasize regions associated with large residual values. The Gaussian parameters are optimized using the Adam optimizer \cite{adam}.

    \item \textbf{Neural network loss minimization:} The optimized Gaussian distribution is then fixed while the neural network parameters $\theta$ are updated to minimize the complete loss given by Eq \ref{eq: L_modified}. Consequently, the neural network focuses its optimization on the regions identified by the residual-based Gaussian tracker. The neural network is initially optimized using Adam \cite{adam}, followed by a transition to the L-BFGS optimizer \cite{LBFGS}.
\end{itemize}

\end{enumerate}

\begin{figure}[h!]
    \centering
    \includegraphics[width=0.9\linewidth]{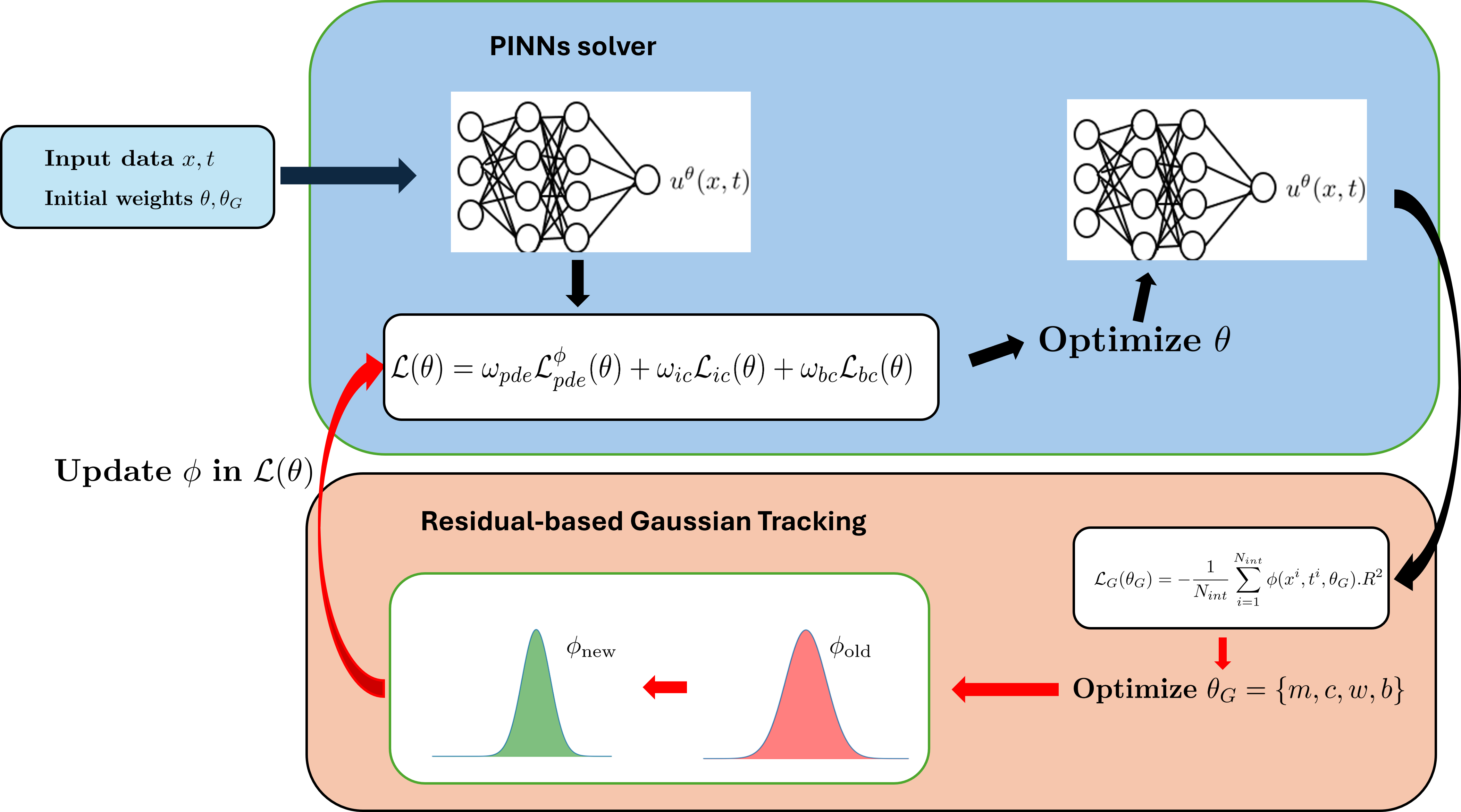}
    \caption{The PINN solver (top) minimizes the residual loss weighted by a dynamic Gaussian distribution $\phi$. Simultaneously, the Residual-based Gaussian Tracking (bottom) performs a linear regression on the PDE residual to center $\phi_{\text{new}}$ in high-error regions. The updated weights are then fed back (red arrow) to focus the PINN's optimization capacity on the moving discontinuity.}
    \label{fig:train_diagram}
\end{figure}

The general workflow of the training strategy  is illustrated in figure \ref{fig:train_diagram}. Algorithm \ref{alg:cap} highlights the training strategy explained above. 
\begin{algorithm}
\caption{\textcolor{blue}{Training algorithm}}\label{alg:cap}
\begin{algorithmic}
  \State \textbf{Warmup phase:} \\
  \State Start the PINN solver with ADAM optimizer to minimize $\mathcal{L}(\theta)$ in equation \ref{eq: loss_standard}
  \Statex
  \State \textbf{ADAM phase:}
  \State \For {$k$ in number of ADAM epochs} 
  \State Run ADAM optimizer for Gaussian tracking to minimize $\mathcal{L}_G(\theta_G)$ in equation \ref{eq: loss_g}
  \State Return $\phi$
  \State Run ADAM optimizer for PINNs to minimize $\mathcal{L}(\theta)$ in equation  \ref{eq: L_modified}
  \State Return $R$ (the PDE residual) 
  \EndFor
  \Statex
  \State \textbf{ADAM L-BFGS phase:}
  \State \For {$k$ in number of ADAM-L-BFGS epochs}
  \State Run ADAM optimizer for Gaussian tracking to minimize $\mathcal{L}_G(\theta_G)$ in equation \ref{eq: loss_g}
  \State Return $\phi$
  \State Run L-BFGS optimizer for PINNs to minimize $\mathcal{L}(\theta)$ in equation  \ref{eq: L_modified}
  \State Return $R$ (the PDE residual) 
  
  \EndFor
  \State  Return $U_{\theta}$
\end{algorithmic}

\end{algorithm}
    
\newpage
\section{Results} \label{sec: results}
The G-PINN methodology is validated with low-viscous Burgers initial-boundary value problems that exhibit both static and moving shock waves. The results obtained are evaluated against the solutions obtained using standard PINN and a Cole-Hopf reference solution; the Cole-Hopf method is highlighted in \ref{appendix: cole-Hopf}. The results presented in this work were obtained using a neural network consisting of $6$ hidden layers, with $20$ neurons each, with a hyperbolic tangent activation function used throughout all layers except the output layer. All predictions presented were conducted using $20000$ interior collocation points and $50$ initial and boundary points. The weights $\omega_{pde}$, $\omega_{ic}$, $\omega_{bc}$  are fixed throughout all test cases with the values $0.3$, $0.7$, and $0.7$, respectively. Notice that throughout this study, we denote as "standard PINNs" the architecture that uses the static weight loss approach defined in section \ref{meth:pinns}. Table \ref{table: Lr} highlights the Learning rate (Lr) values used to obtain the results of 1D low-viscous Burgers presented in the work. 

\begin{table}[h!]
    \centering
   \begin{tabular}{c|c|c|c}
   \hline
    Case & Lr ADAM (Gaussian tracking)  & Lr ADAM & Lr L-BFGS  \\
    \hline
    standard PINN & - & $10^{-3}$ & $4 \times 10^{-2}$\\
    G-PINN & $10^{-2}$ & $10^{-3}$& $4 \times 10^{-2}$ \\
    \hline
\end{tabular}
   \caption{Learning rates for G-PINN and standard PINN}
    \label{table: Lr}
\end{table}

\subsection{Static shock wave}\label{burgers: 0.0005-static} \label{subsec: static _shock_0005}
\begin{figure}[h!]
    \centering
    \begin{subfigure}
        {0.45\textwidth}
         \centering
         \includegraphics[width=\textwidth]{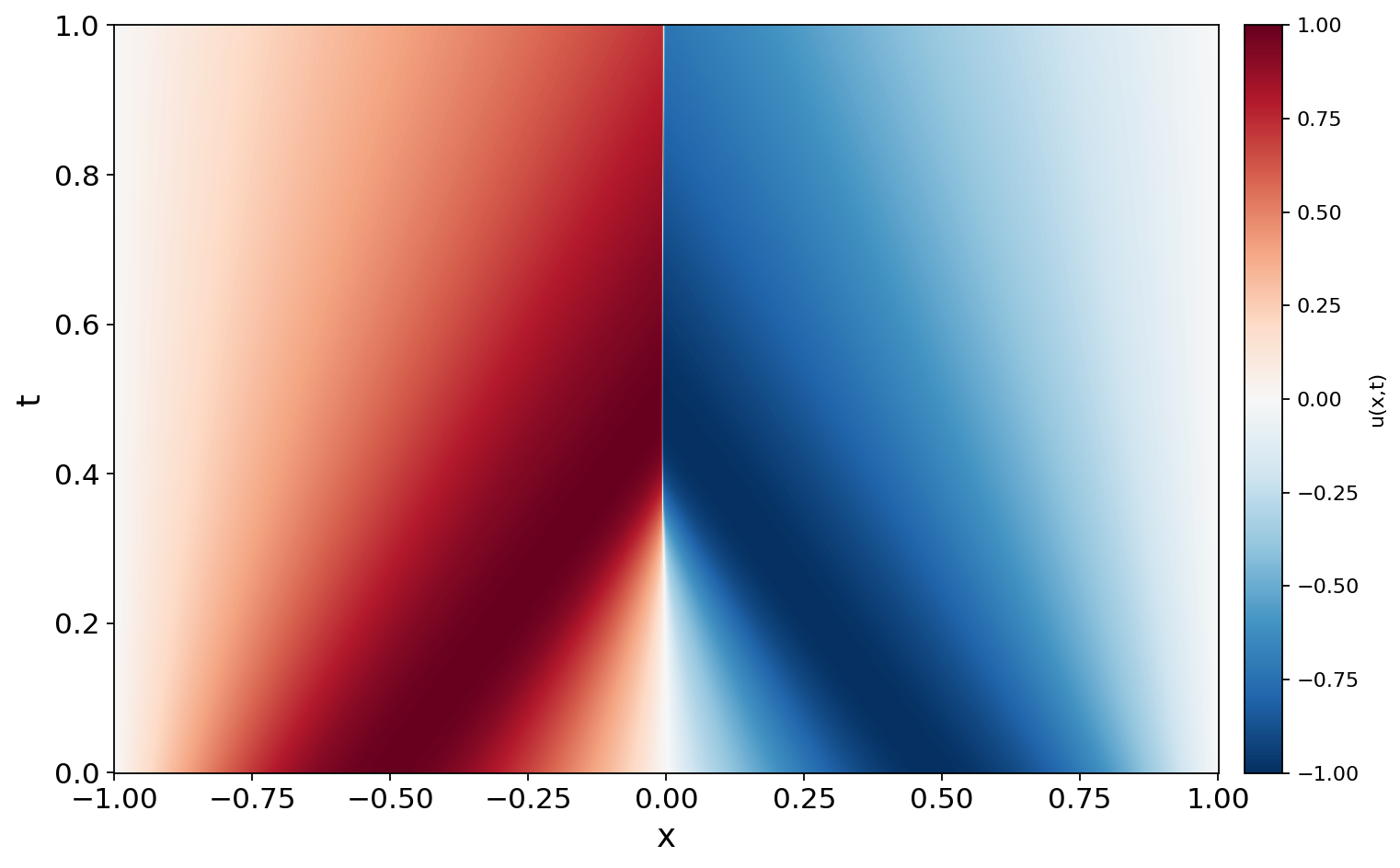}
         \caption{ }
         \label{fig:nu_0005_gmm_pinn}
    \end{subfigure}
    \begin{subfigure}
        {0.45\textwidth}
         \centering
         \includegraphics[width=\textwidth]{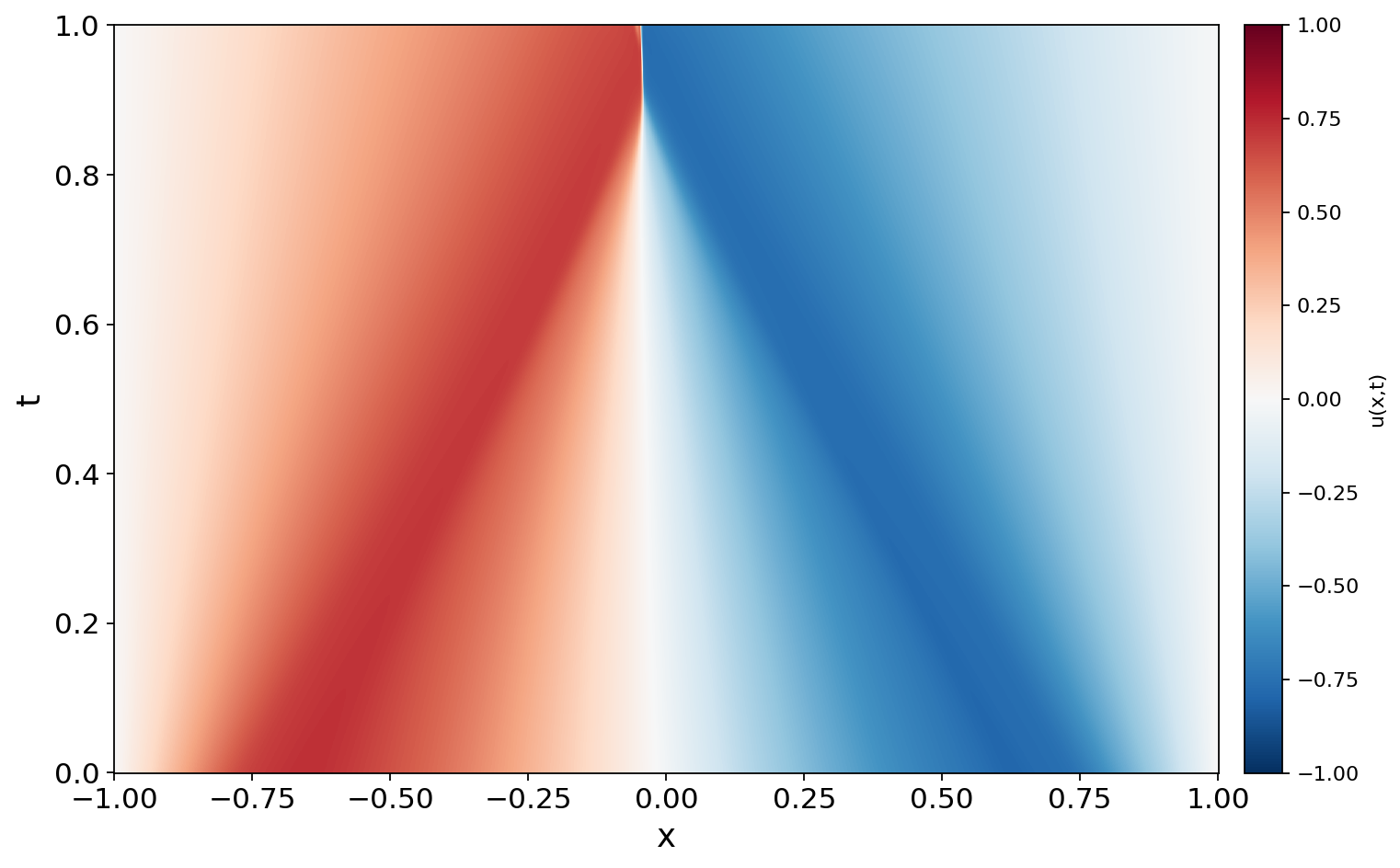}
         \caption{ }
         \label{fig:nu_0005_pinn}
    \end{subfigure}
    \begin{subfigure}
        {0.5\textwidth}
         \centering
         \includegraphics[width=\textwidth]{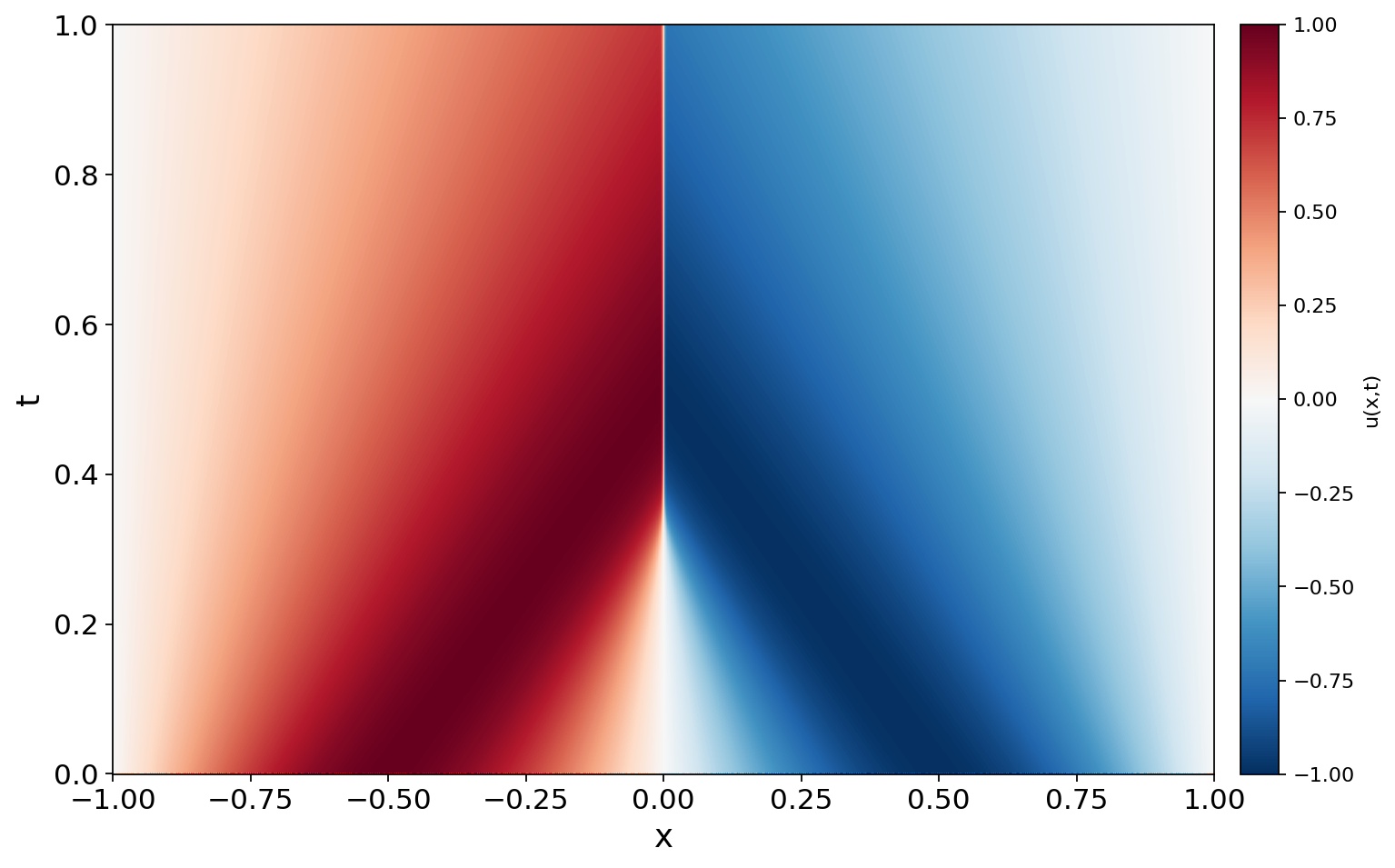}
         \caption{ }
         \label{fig:nu_0005_ref}
    \end{subfigure}
    \caption{Solutions obtained using G-PINNs: Gaussian-based spatially weighted formulation \ref{fig:nu_0005_gmm_pinn}, standard PINN \ref{fig:nu_0005_pinn} and reference solution obtained with Cole-Hopf transformation \ref{fig:nu_0005_ref}.}
    \label{fig: solution_nu_0005}
\end{figure}

To assess the capability of the  G-PINNs methodology, we first consider a low viscosity in Burgers' equation.
$$
\begin{cases}
    u_t + uu_x = 0.0005 u_{xx},  \quad x\in [-1,1], \quad t\in [0,1]\\
    u(x,0) = -\sin(\pi x) \\
     u(-1,t) = u(1,t) = 0.
\end{cases}
$$
This case exhibits a sharper shock wave at $x=0$ with thickness $o(\nu)$, compared to the previous test case.  Figure \ref{fig: solution_nu_0005} presents the solutions of the proposed model, the standard PINN and the Cole-Hopf 
reference solution. The findings emphasize a close agreement between the solution obtained through the proposed methodology and the reference solution; see figures \ref{fig:nu_0005_gmm_pinn} and \ref{fig:nu_0005_ref}. In contrast, standard PINN fails for this case as can be seen in figure \ref{fig:nu_0005_pinn}, confirming the inherent difficulties that standard PINN face in low-viscosity regimes. 

\begin{figure}[h!]
    \centering
    \includegraphics[width=0.9\linewidth]{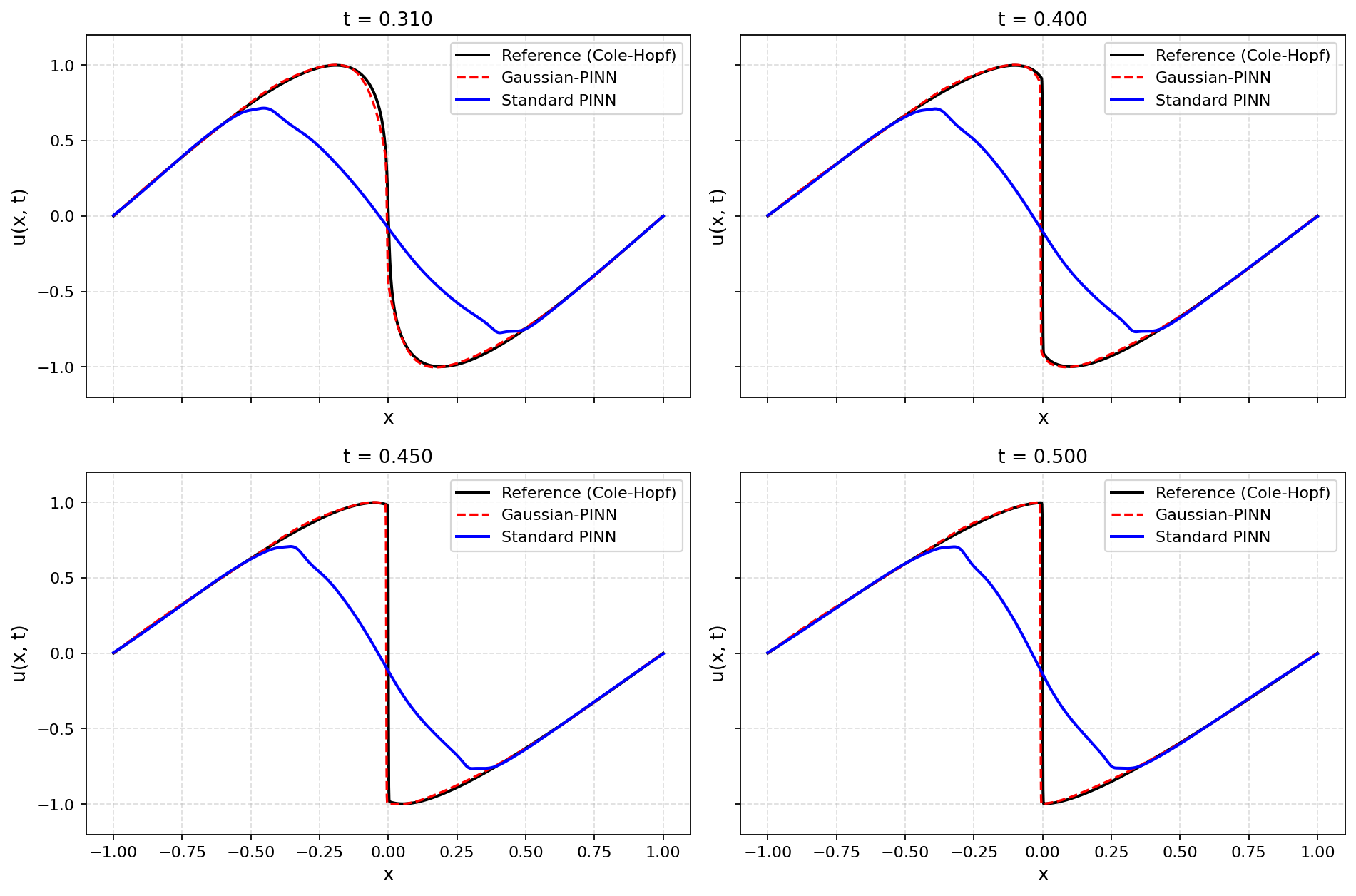}
    \caption{Solutions profiles at different time instants.}
    \label{fig:nu_0005_slices}
\end{figure}

Figure \ref{fig:nu_0005_slices} shows the solution at various time instants. The proposed model matches almost perfectly the reference solution, In particular at $t=0.31$ which corresponds to the neighborhood of the breaking time. The standard PINN solution drastically deviates from the reference solution, confirming that standard PINN fails to resolve the shock wave at this low-viscosity regime.
\begin{figure}[h!]
    \centering
    \begin{subfigure}{0.49\textwidth}
    \includegraphics[width=\linewidth]{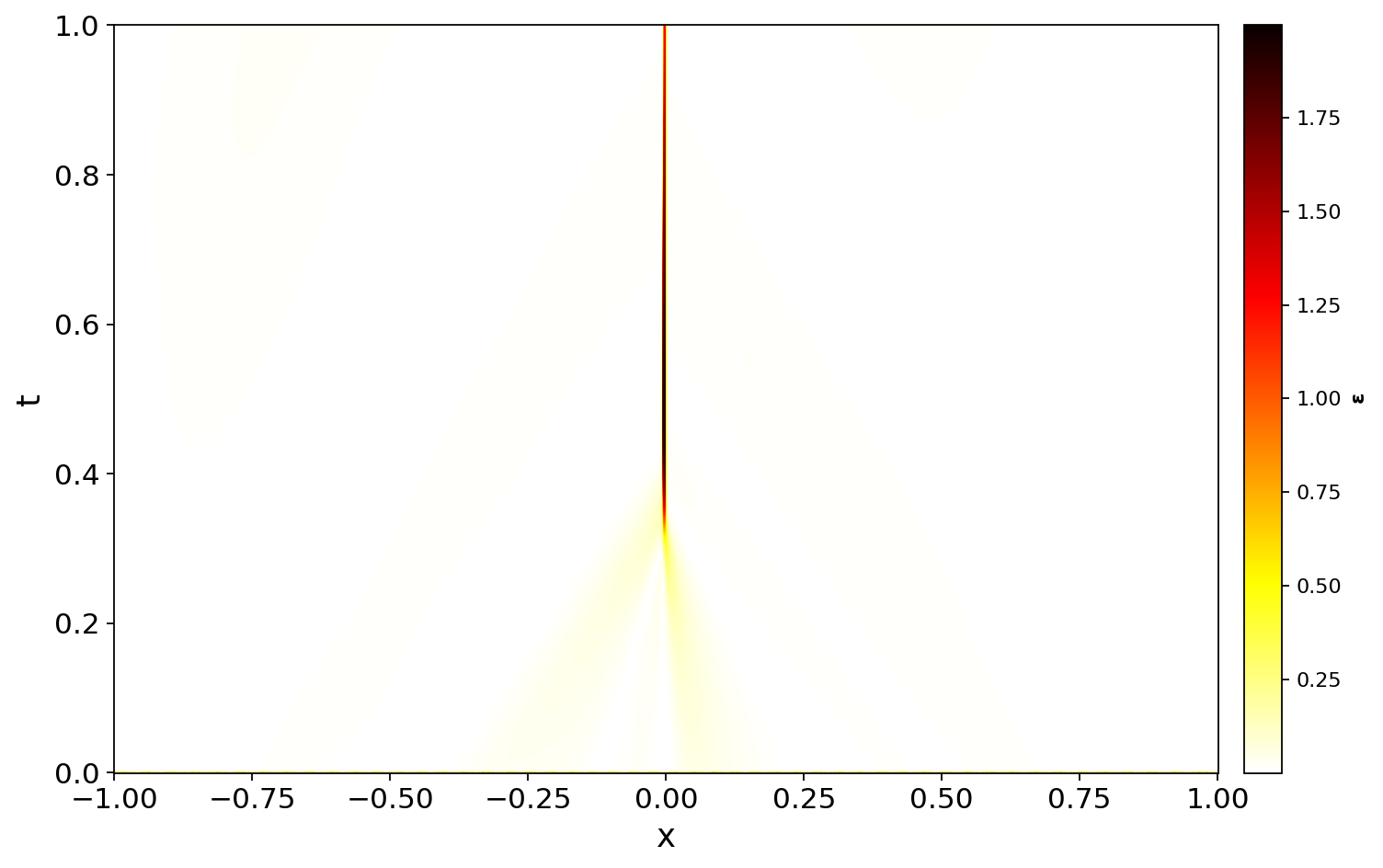}
    \caption{ }
    \label{fig: nu_0005_error_gmm}    
    \end{subfigure}
        \begin{subfigure}{0.49\textwidth}
    \includegraphics[width=\linewidth]{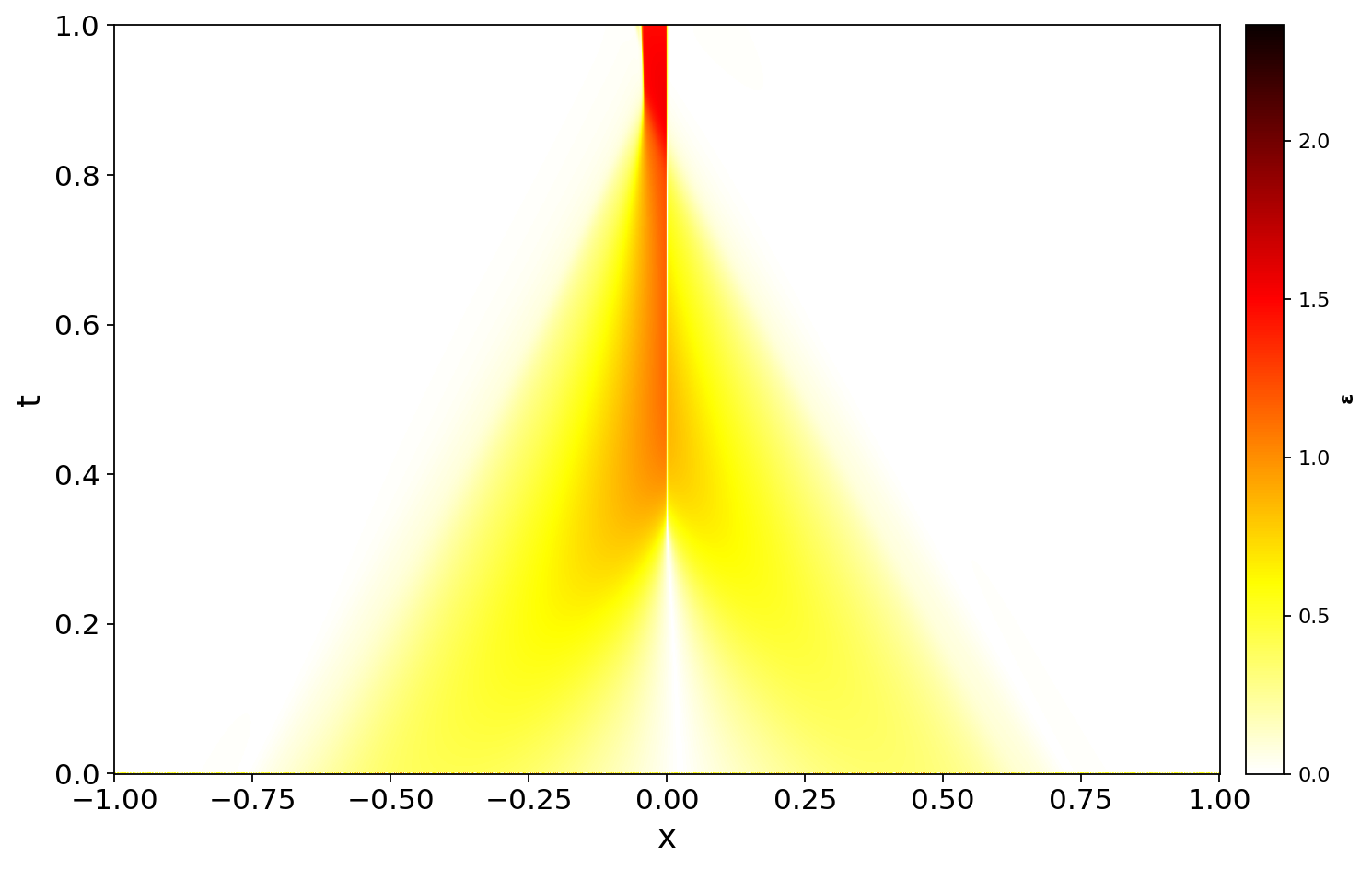}
    \caption{ }
    \label{fig: nu_0005_error_pinn}    
    \end{subfigure}
    \caption{Absolute error maps of the G-PINNs: Gaussian-based spatially weighted loss formulation method \ref{fig: nu_0005_error_gmm} and the standard PINN \ref{fig: nu_0005_error_pinn}.} 
    \label{fig: nu_0005_error}
\end{figure}
In figure \ref{fig: nu_0005_error}, we show the absolute error maps obtained for the proposed method and for standard PINNs. 

Figure~\ref{fig: nu_0005_error_gmm} shows the absolute error of the proposed method. The error is spatially concentrated along a sharp vertical line at $x=0$, where the shock wave forms, and reaches its maximum around $t=0.3$ -- $0.4$. The errors decrease as time progresses (dropping from 1.75-1.8 (dark red) to 0.75-1 (light red-orange)), confirming that the Gaussian linear regression model can track the shock wave in time, allowing the model to capture the shock as time evolves. Figure~\ref{fig: nu_0005_error_pinn} shows the error obtained with the standard PINN. The error is large around the shock wave and increases with time, supporting the results shown in Figures~\ref{fig:nu_0005_pinn} and~\ref{fig:nu_0005_slices}.
\begin{table}[h!]
    \centering
    \begin{tabular}{c|c|c}
    \hline
    Case & L2& MAE\\  
    \hline
    standard PINN & $4.498\times 10^{-1}$ & $1.332\times 10^{-1}$ \\
    G-PINN & $1.335\times 10^{-1}$ & $1.205\times 10^{-2}$ \\
    \hline
\end{tabular} 
    \caption{Viscous Burgers' Eq. with $\nu = 0.0005$: Error metrics}
    \label{table: nu_0005}
\end{table}

Table \ref{table: nu_0005} summarizes the L2 error and MAE for the G-PINN formulation and the standard PINNs, The results reveal a relative L2 error of approximately $45\%$ for standard PINN and only $13.3 \%$ for the proposed model, this superiority is also confirmed by the values of MAE, namely $1.332\times 10^{-1}$ for standard PINN compared to $1.205\times 10^{-2}$ for the methodology introduced. 

This case demonstrates the capability of the proposed framework to track and resolve a stationary shock wave. By eliminating the need for \textit{a priori} knowledge of the shock position or localized mesh refinement, the proposed method achieves high fidelity and improved generalization in capturing steep-gradient regimes.
\newpage
\subsection{Moving shock wave} \label{subsec: moving _shock_0005}
This section is dedicated to highlighting the results obtained for a moving shock wave solution of the following initial-boundary value problem:   

$$
\begin{cases}
    u_t + uu_x = 0.0005u_{xx} \\
    u(x,0) = -\sin(\pi x) + 0.50 \\
    u(-1,t) = u^*(-1,t), \quad u(1,t) = u^*(1,t),
\end{cases}
$$
where $u^*$ is the Cole-Hopf reference solution.

\begin{figure}[h!]
    \centering
    \begin{subfigure}
        {0.45\textwidth}
         \centering
         \includegraphics[width=\textwidth]{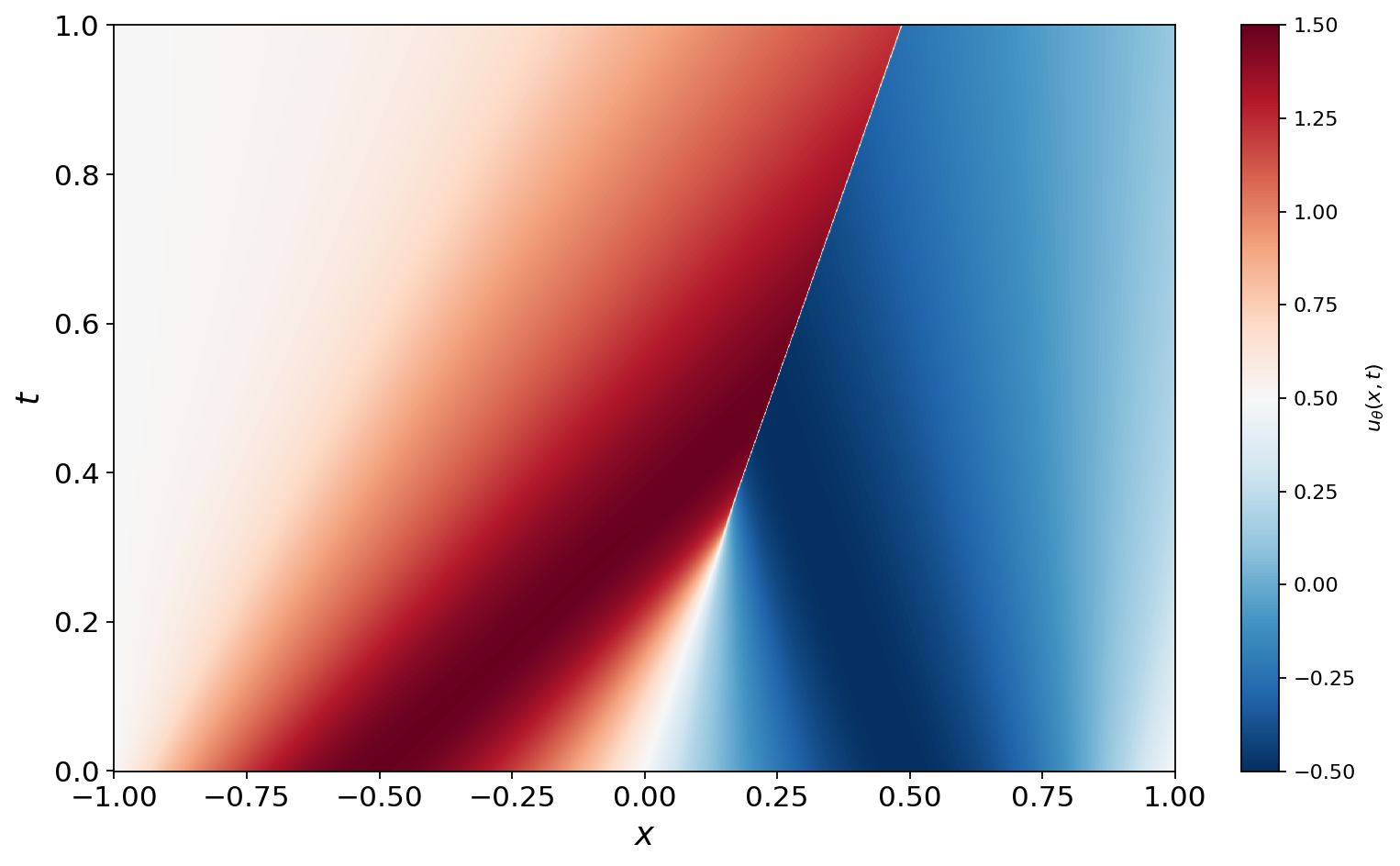}
         \caption{ }
         \label{fig:nu_0005_moving_gmm_pinn}
    \end{subfigure}
    \begin{subfigure}
        {0.45\textwidth}
         \centering
         \includegraphics[width=\textwidth]{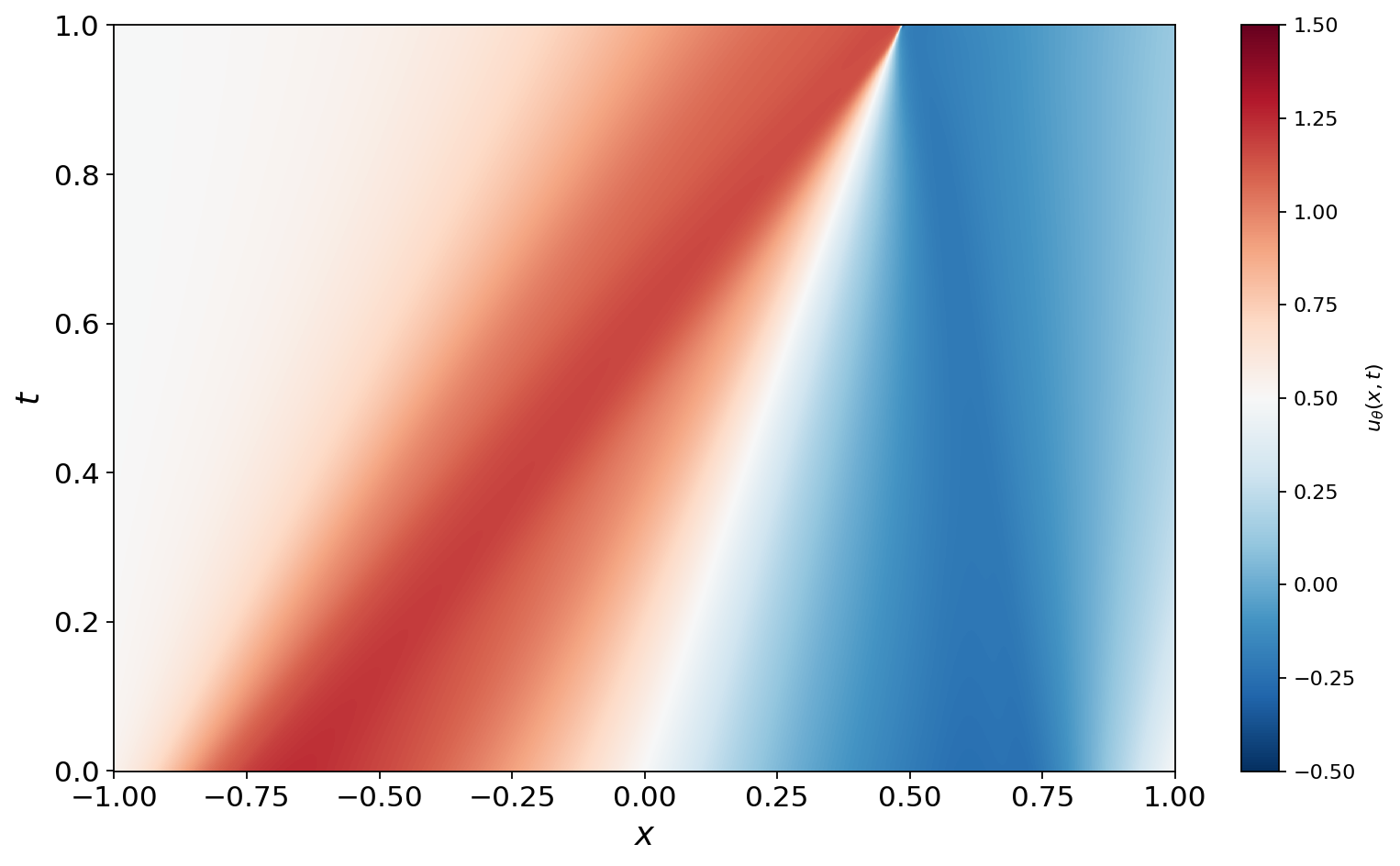}
         \caption{ }
         \label{fig:nu_0005_moving_pinn}
    \end{subfigure}
    \begin{subfigure}
        {0.5\textwidth}
         \centering
         \includegraphics[width=\textwidth]{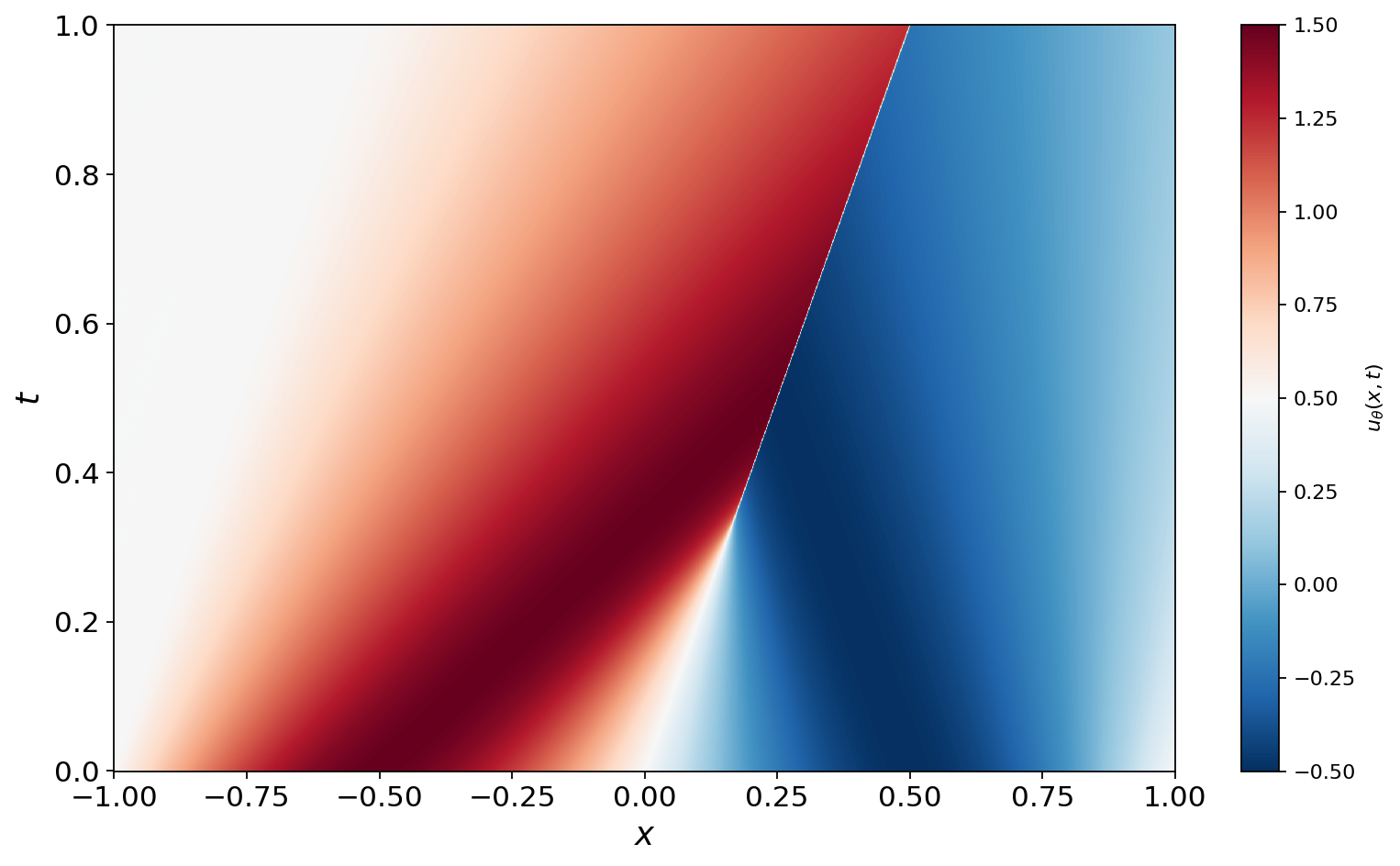}
         \caption{ }
         \label{fig:nu_0005_moving_ref}
    \end{subfigure}
    \caption{Solutions obtained using G-PINNs: Gaussian-based spatially weighted formulation \ref{fig:nu_0005_moving_gmm_pinn}, standard PINN \ref{fig:nu_0005_moving_pinn} and reference solution obtained with Cole-Hopf transformation \ref{fig:nu_0005_moving_ref}.}
    \label{fig: solution_nu_moving_0005}
\end{figure}

\begin{figure}
    \centering
    \includegraphics[width=1.0\linewidth]{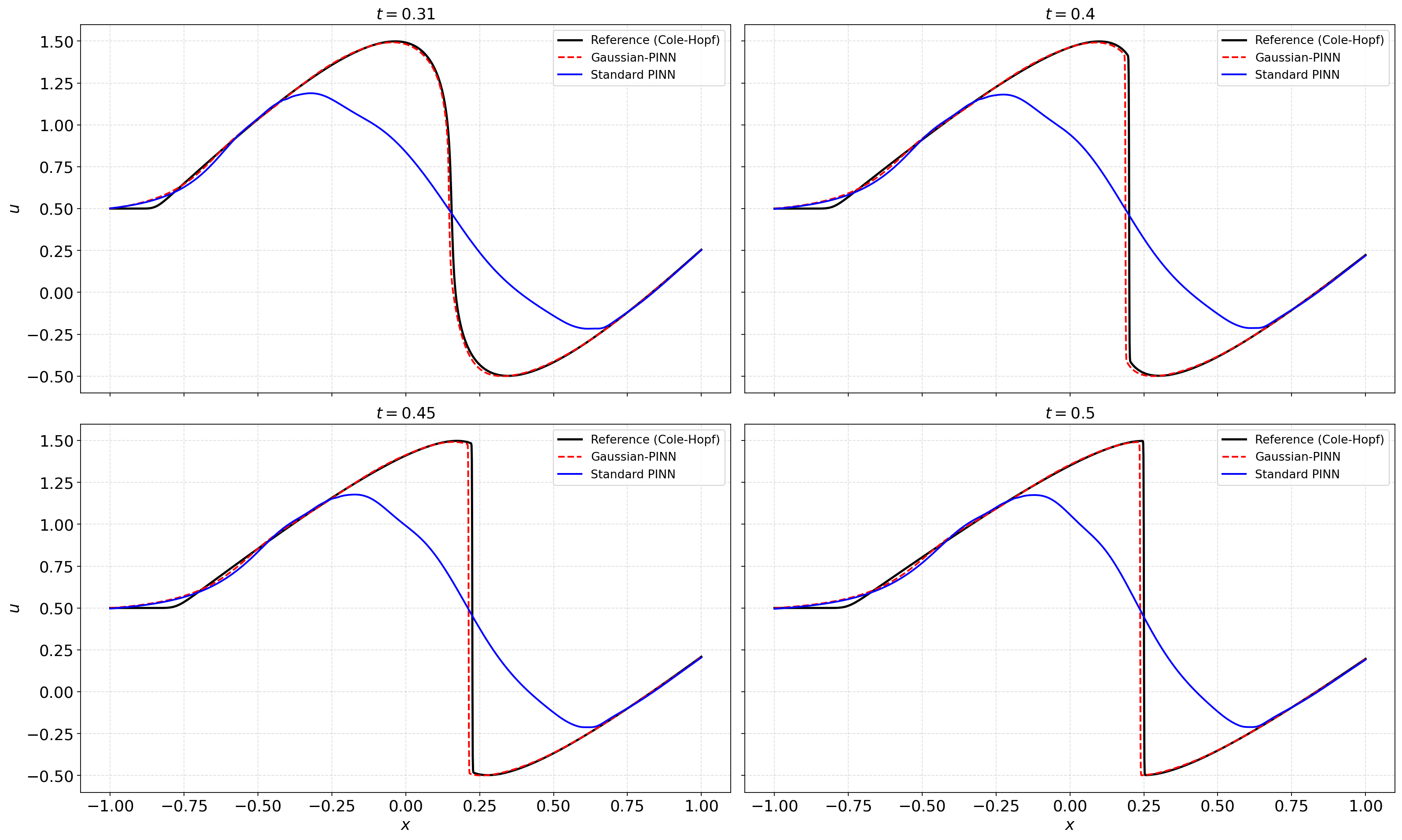}
    \caption{Solutions profiles at different time instants.}
    \label{fig:nu_0005_moving_slices}
\end{figure}

This case is selected to challenge the method's ability to track a moving shock wave. Unlike for the static case \ref{burgers: 0.0005-static}, now the initial profile of this case is shifted by a constant $0.5$, resulting in a shock propagating to the right that traverses the domain and provides a rigorous test for the proposed methodology. Figure \ref{fig: solution_nu_moving_0005} shows the results obtained for the proposed model, the standard PINNs, and the reference Cole-Hopf transformation. The solution of the proposed method shows close visual agreement with the reference solution. The standard PINN clearly failed to resolve the shock wave; see figure \ref{fig:nu_0005_moving_pinn}. 
The results show that the proposed method could effectively predict the break time ($t^* \approx 0.31-0.33$), which is consistent with the theoretical break time ($t = \frac{1}{\pi}$). Moreover, the method could also predict the shock layer (the learned shock wave speed $m \approx 0.45-0.49$), see figure \ref{fig:nu_0005_moving_gmm_pinn}. This prediction is in alignment with the reference shock wave speed for this case ($m^* = 0.5$). Similarly to the previous test cases, the snapshots of the solution in $ t \in \{ 0.31, 0.40, 0.45, 0.50\}$ are presented in figure \ref{fig:nu_0005_moving_slices}, the formulation of G-PINN could effectively resolve the propagating shock wave at different time instants  following the breaking shock time, demonstrating that the proposed is capable of tracking and capturing the shock wave effectively. The standard PINN cannot resolve the shock wave in this case during the time interval following shock formation ($t \geq \frac{1}{\pi}$)

\begin{figure}[h!]
    \centering
    \begin{subfigure}{0.49\textwidth}
    \includegraphics[width=\linewidth]{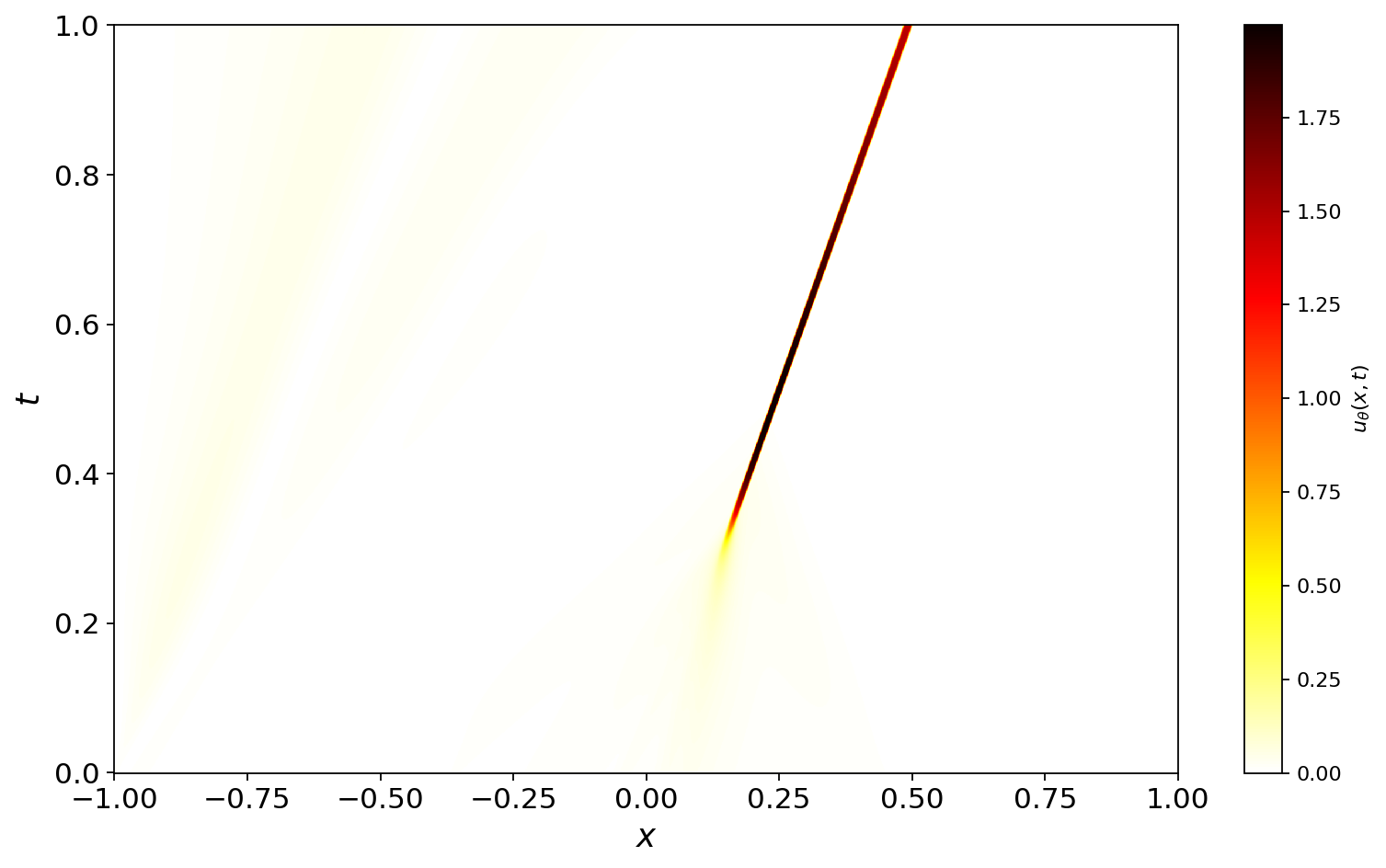}
    \caption{ }
    \label{fig: nu_0005_moving_error_gmm}    
    \end{subfigure}
        \begin{subfigure}{0.49\textwidth}
    \includegraphics[width=\linewidth]{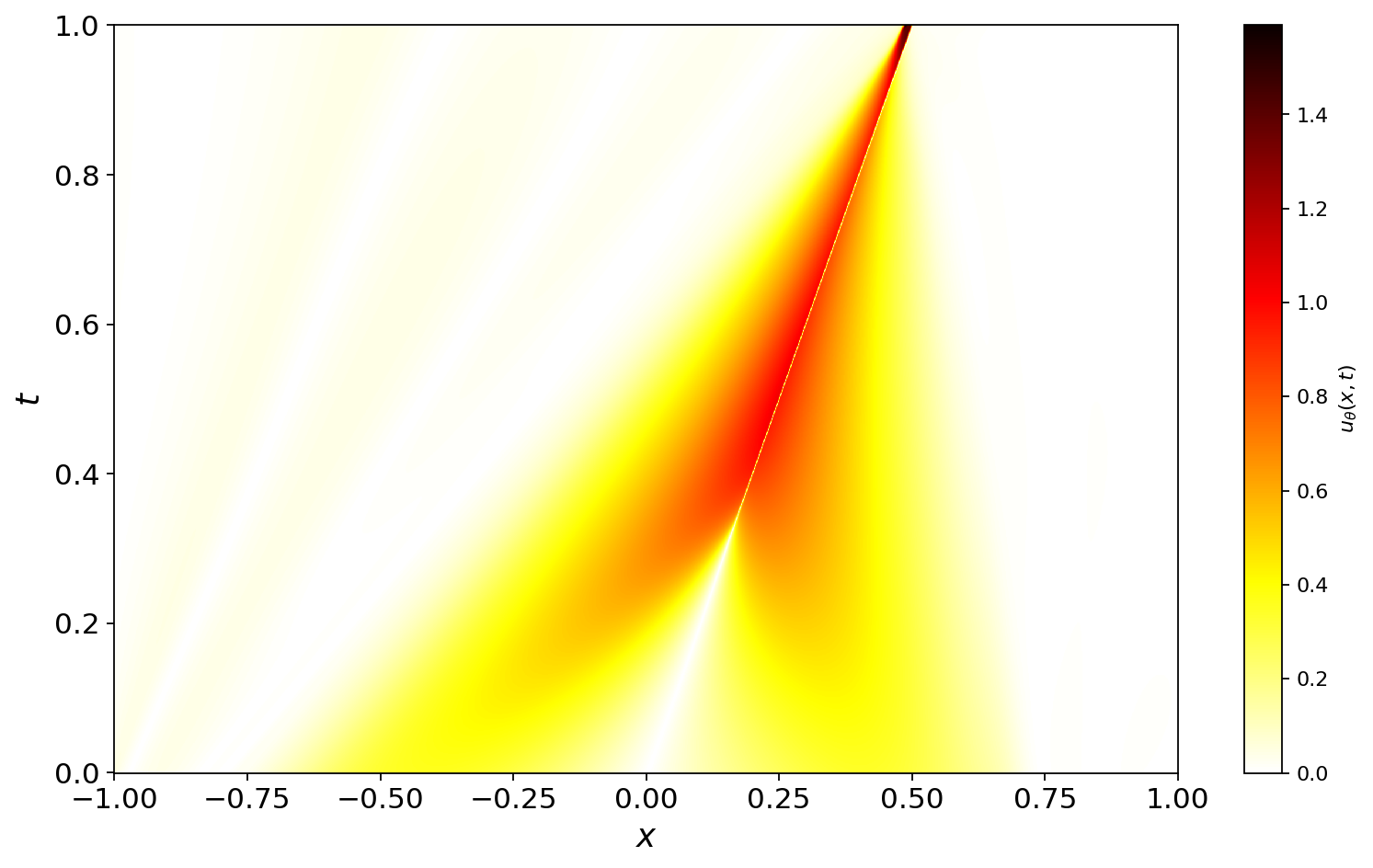}
    \caption{ }
    \label{fig: nu_0005_moving_error_pinn}    
    \end{subfigure}
    \caption{Absolute error maps of the G-PINNs: Gaussian-based spatially weighted loss formulation method \ref{fig: nu_0005_moving_error_gmm} and the standard PINN \ref{fig: nu_0005_moving_error_pinn}.}
    \label{fig: nu_0005_moving_error}
\end{figure}

Figure~\ref{fig: nu_0005_moving_error} shows the absolute error maps for this test case. The proposed methodology confines the error to a sharp line aligned with the trajectory of the shock-wave. The error progressively decreases in intensity as time evolves ($t \geq 0.4$), reflecting the increasing fidelity of the model to capture the propagating shock wave. This behavior is also consistent with the fact that the early post-breaking phase involves rapid shock steepening, which constitutes the most challenging prediction stage for the neural network. At later times, well beyond the break time, the shock wave becomes more stable and propagates at an approximately constant speed. In contrast, Figure~\ref{fig: nu_0005_moving_error_pinn} shows that, for the standard PINN, the absolute error spreads well beyond the shock wave, demonstrating that the standard PINN fails to localize the discontinuity in this test case.

The L2 and MAE values for this case are reported in table \ref{table: nu_0005_moving}. The standard PINN yield a $32.2\%$ L2 error and $1.44 \times 10^{-1}$ MAE, compared to only $14.4\%$ L2 and $6.69 \times 10^{-2}$ MAE for the proposed methodology. 

\begin{table}[h!]
    \centering
    \begin{tabular}{c|c|c}
    \hline
    Case & L2& MAE\\  
    \hline
    standard PINN & $3.252 \times 10^{-1}$ & $1.447\times 10^{-1}$ \\
    G-PINN & $1.441\times 10^{-1}$ & $6.694\times 10^{-2}$ \\
    \hline
\end{tabular} 
    \caption{Viscous Burgers' Eq. with $\nu = 0.0005$, moving shock wave: Error metrics.}
    \label{table: nu_0005_moving}
\end{table}

The proposed method successfully tracked the moving shock wave without requiring a priori knowledge of its trajectory, providing a high-fidelity solution without the need to manually refine the spatio-temporal domain.

\section{Two dimensional numerical experiments: Riemann Problem}
\begin{figure}[h!]
    \centering
    \begin{subfigure}
        {0.45\textwidth}
         \centering
         \includegraphics[width=\textwidth]{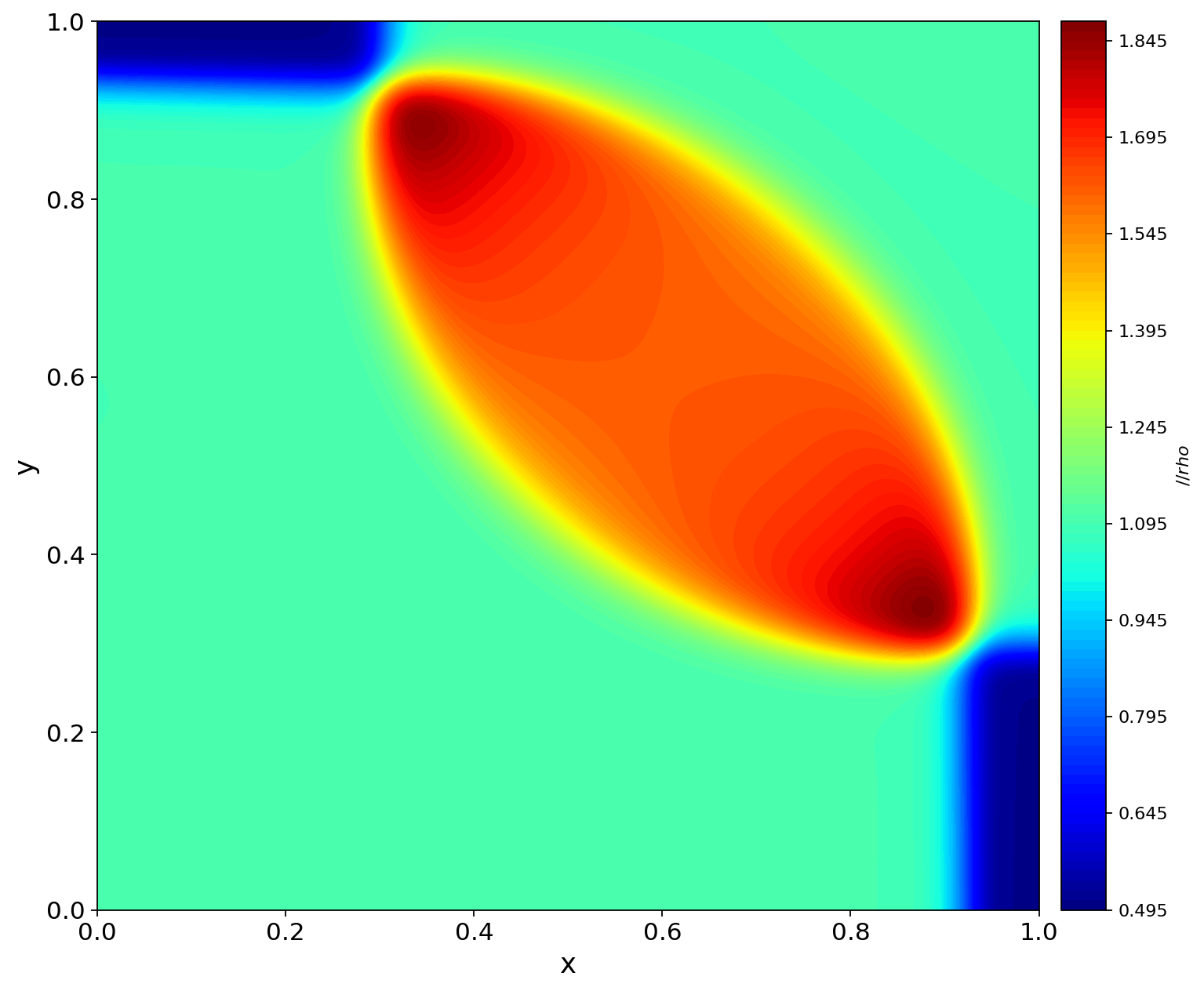}
         \caption{ }
         \label{fig:2d-gpinn_rho}
    \end{subfigure}
    \begin{subfigure}
        {0.45\textwidth}
         \centering
         \includegraphics[width=\textwidth]{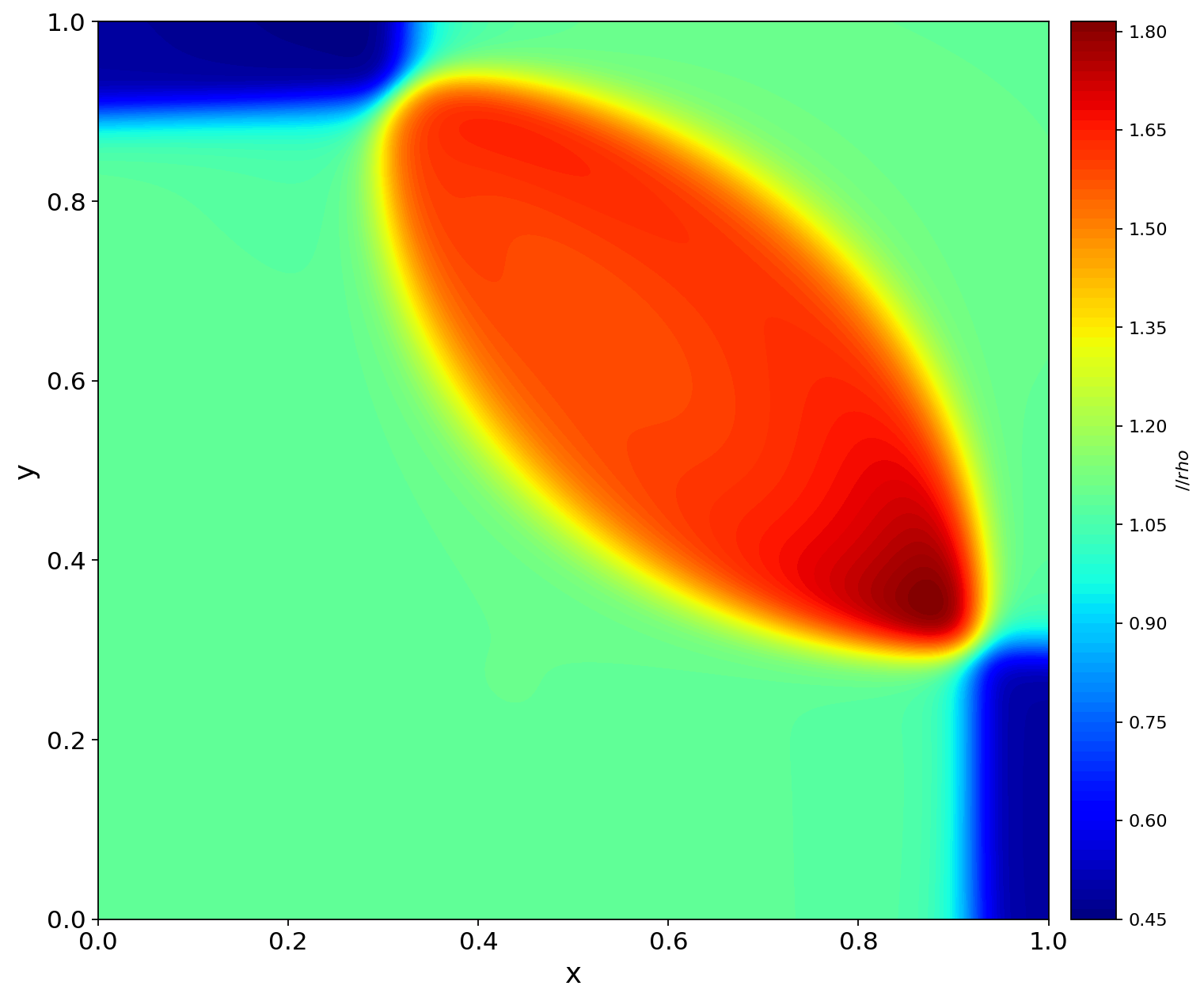}
         \caption{ }
         \label{fig:2d-pinn_rho}
    \end{subfigure}
    \begin{subfigure}
        {0.5\textwidth}
         \centering
         \includegraphics[width=\textwidth]{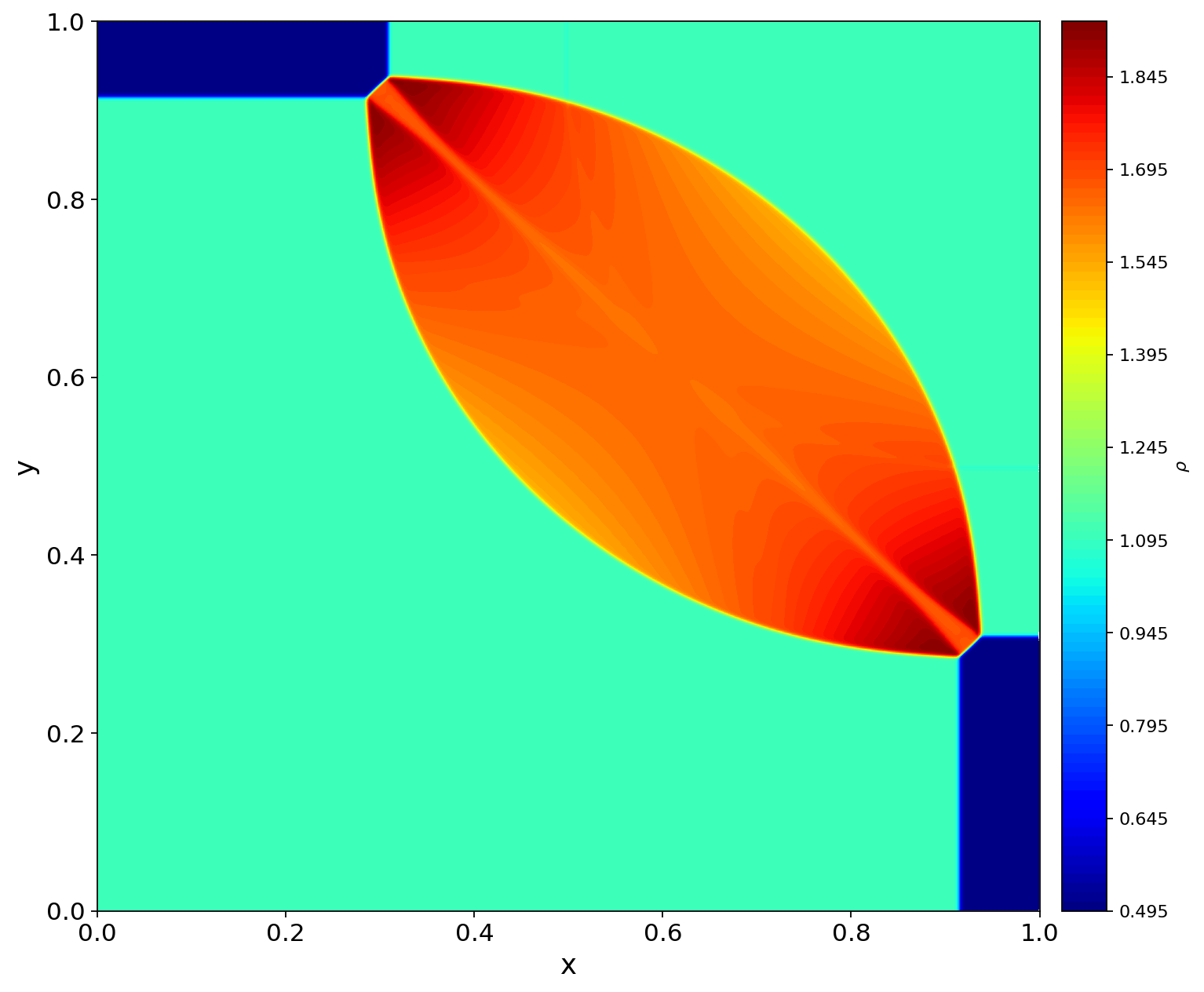}
         \caption{ }
         \label{fig:2d-cfd_rho}
    \end{subfigure}
    \caption{Density contours obtained using G-PINNs: Gaussian-based spatially weighted formulation \ref{fig:2d-gpinn_rho}, standard PINN \ref{fig:2d-pinn_rho} and CFD \ref{fig:2d-cfd_rho}}
    \label{fig: 2d-rho}
\end{figure}

\begin{figure}[h!]
    \centering
    \begin{subfigure}{0.49\textwidth}
    \includegraphics[width=\linewidth]{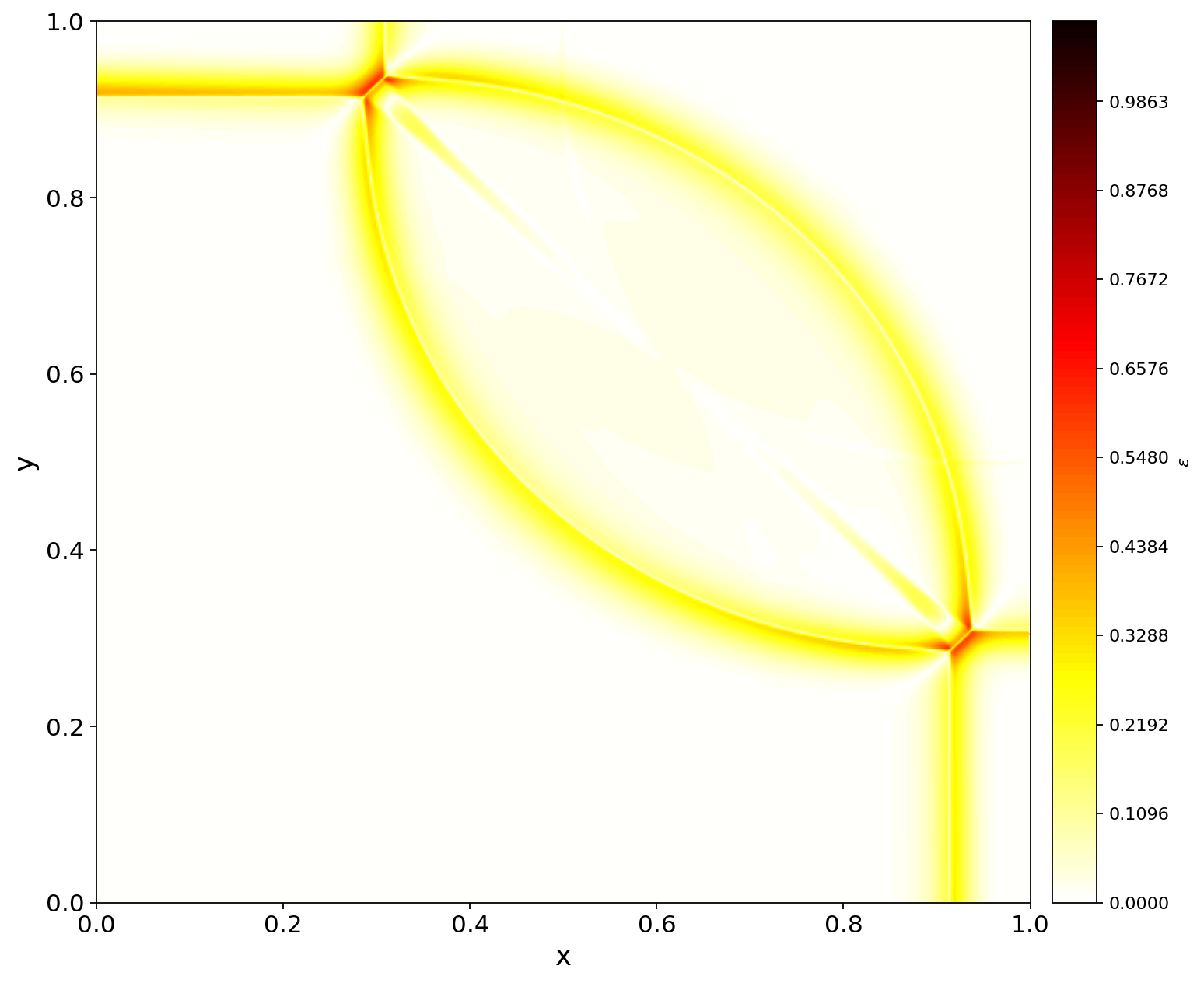}
    \caption{ }
    \label{fig: 2d-gpinn_error}    
    \end{subfigure}
        \begin{subfigure}{0.49\textwidth}
    \includegraphics[width=\linewidth]{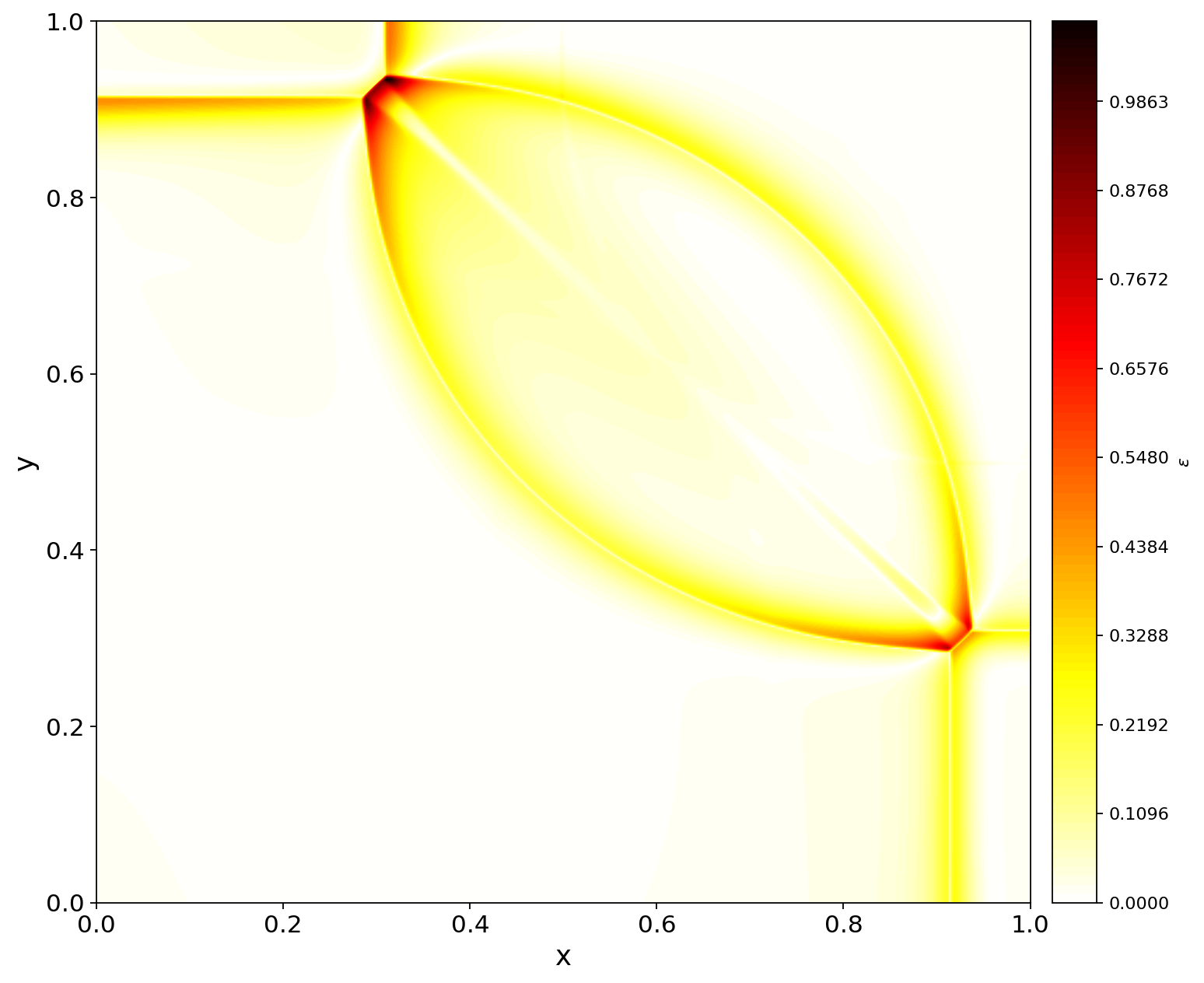}
    \caption{ }
    \label{fig: 2d-pinn_error}    
    \end{subfigure}
    \caption{Absolute error maps of the G-PINNs: Gaussian-based spatially weighted loss formulation method \ref{fig: 2d-gpinn_error} and the standard PINN \ref{fig: 2d-pinn_error}}.
    \label{fig: 2d_error}
\end{figure}

Having demonstrated the potential of G-PINNs in one-dimensional problems, we now assess its capabilities for a 2D Euler Riemann problem. The 2D space and time computational domain for the problem considered is $\{ (x,y,t) \in [0,1] \times [0,1] \times [0, t_{end}]\}$. 
The domain is divided into 4 quadrants, each initialized with a constant state so that neighboring states are connected by a shock wave, slip line, or rarefaction. These initial states are determined by different configurations; see \cite{2dRiemann_2cfg} and \cite{litte2D_1}. In this study, we focus on Configuration 3 see details in table \ref{table: 2D-config}. 
\begin{table}[h]
\centering
\caption{Initial conditions for the 2D Riemann Problems.}
\label{tab:riemann_ic}
\renewcommand{\arraystretch}{1.3}
\begin{tabular}{c|cccc}
\hline
\multirow{2}{*}{Region} 
& \multicolumn{4}{c}{Configuration 3} 
\\
& $\rho$ & $u$ & $v$ & $P$ 
 \\
\hline
$1$ & 1.1    & 0      & 0      & 1.1  \\
2 &  0.5065 & 0.8939 & 0      & 0.35 \\
3  & 1.1    & 0.8939 & 0.8939 & 1.1  \\
4 &  0.5065 & 0      & 0.8939 & 0.35 \\
\hline
\end{tabular}
\label{table: 2D-config}
\end{table}
The initial states evolve into $4$ interacting shock waves that propagate through the domain until the final time $t_{\text{end}} = 0.4$ is reached. To dynamically track the four shock waves, we propose using a mixture of 4 Gaussian distributions $\Phi$:
\begin{equation}
    \Phi = \sum_{i=1}^{4} \kappa_{i}\phi_{i}
\end{equation}
where $\kappa_{i}$ represents the mixing coefficient (or weight) for each component that satisfies $\sum \kappa_i = 1$, and $\phi_{i}$ is a two-dimensional Gaussian distribution defined as follows:
\begin{equation}
\phi_{i}(x, y, t) = \frac{1}{2\pi \sigma_{x,i}(t) \sigma_{y,i}(t)} e^{ \left( -\frac{1}{2} \left[ \left(\frac{x - \mu_{x,i}(t)}{\sigma_{x,i}(t)}\right)^2 + \left(\frac{y - \mu_{y,i}(t)}{\sigma_{y,i}(t)}\right)^2 \right] \right)}
\end{equation}
where $\mu_{x,i}(t)$ and $\mu_{y,i}(t)$ represent the spatial centers along the the axes $x$ and $y$ defined as follows:
    \begin{align*}
        \mu_{x,i}(t) = m_{x,i}t+c_{x,i} \\
        \mu_{y,i}(t) = m_{y,i}t+c_{y,i},
    \end{align*}
    and $\sigma_{x,i}(t)$ and $ \sigma_{y,i}(t)$ are the standard deviations along the $x$ and $y$ axes:
    \begin{align*}
    \sigma_{x,i}(t)
    &= \ln\left(1+e^{w_{x,i}t+b_{x,i}}\right),\\
    \sigma_{y,i}(t)
    &= \ln\left(1+e^{w_{y,i}t+b_{y,i}}\right).
\end{align*}
The formulations of $\mu_{x,i}$, $\mu_{x,i}$, $\sigma_{x,i}$ and $ \sigma_{y,i}$ are natural extensions of the 1D formulation presented in \ref{subsec: G-pinn formulation}. The set of learnable parameters for the Residual-based Gaussian tracking $\theta_G$ becomes:
$$
\theta_G = \{ m_x, m_y, c_x, c_y, w_x, w_y, b_x, b_y, \kappa \}
$$
Notice that $\kappa$ (the mixing coefficient controlling the four Gaussian distributions) is also included as a learnable parameter to determine the contribution of each Gaussian component in the multi-modal Gaussian $\Phi$. For this  case, we used a wider neural network compared to the 1D cases, with 6 hidden layers each with 48 neurons. The training follows the same process as described in the algorithm \ref{alg:cap}. Figure \ref{fig: 2d-rho} represents the density contours $\rho$ obtained using G-PINN, standard PINN, and CFD, the CFD solution is simulated using PyClaw, a hyperbolic PDEs solver, see \cite{clawpack, pyclaw-sisc}. The CFD solution is obtained using a $400 \times 400$ mesh. The density contours reveal that both G-PINN and standard PINN could capture the overall structure of the four shock waves, including the low-density near the boundaries, although the solutions obtained by G-PINN and standard PINN are slightly diffusive compared to the CFD solution, it is worth noting that this test case is highly challenging, as pointed out by \cite{compressible2dshock-pinns}. The G-PINN solution shows better agreement with the CFD result, reaching a peak density closer to the reference value, see figures \ref{fig:2d-gpinn_rho} and \ref{fig:2d-cfd_rho}. 
Figure \ref{fig: 2d_error} shows the absolute error $\epsilon$ for G-PINNs and standard PINNs. The G-PINN error shows a narrower and less intense error band along the shock curves  than the standard PINN, this is also observed in the low-density corners. In addition, the maximum level error of the standard PINN reached $1.08$ compared to $0.62$ for the G-PINN. This indicates that G-PINN could localize the discontinuities more sharply and reduces the overall error magnitude, which supports the improved accuracy of the G-PINN approach over the standard PINN for this configuration.
\section{Conclusion} \label{sec: conclusion}
This study introduced a residual-tracked Gaussian weighting strategy to improve the representation of shocks and other localized high-gradient features with physics-informed neural networks. Rather than changing the collocation distribution, the method modifies the PDE residual loss through a low-dimensional Gaussian weighting whose parameters are updated from the evolving residual field. This alternating optimization concentrates training effort near the most difficult region of the solution and allows the weighting field to follow stationary and moving shocks without prescribing their locations or trajectories.

For the one-dimensional Burgers benchmarks, the proposed method consistently reduced the error relative to the standard PINN, with the largest improvements obtained at the lower viscosity. At $\nu=5\times10^{-4}$, the relative $L_2$ error decreased from approximately $45\%$ to $13.3\%$ for the stationary shock and from approximately $32.5\%$ to $14.4\%$ for the moving shock. In the moving case, the learned Gaussian trajectory also recovered a shock speed close to the reference value. For the two-dimensional Euler Riemann problem, the Gaussian-mixture extension produced sharper interacting discontinuities and reduced the maximum absolute density error from $1.08$ to $0.62$ relative to the standard PINN.

These results show that residual-tracked Gaussian weighting provides a compact mechanism for concentrating PINN optimization in shock-dominated problems while retaining the original collocation set. The present benchmarks are intended to isolate the effect of the proposed weighting relative to a standard PINN baseline. Future work will extend the assessment to systematic comparisons with adaptive sampling and adaptive weighting strategies, robustness studies over multiple network initializations, and accuracy--cost analyses for larger multidimensional compressible-flow problems. More general Gaussian-mixture parameterizations will also be investigated for curved, interacting, and topologically changing shock structures.

\section*{Acknowledgments}
Esteban Ferrer acknowledge the funding from the European Union (ERC, Off-coustics, project number 101086075). Views and opinions expressed are, however, those of the authors only and do not necessarily reflect those of the European Union or the European Research Council. Neither the European Union nor the granting authority can be held responsible for them.
Esteban Ferrer acknowledges the funding received by the Grant DeepCFD (Project No. PID2022-137899OB-I00) funded by MICIU/AEI/10.13039/501100011033 and by ERDF, EU.
\section*{Data Availability}
The data that support the findings of this study are available from the corresponding author upon reasonable request.
\newpage

\appendix
\section{Additional diagnostics for the stationary shock wave Burgers case }
\begin{figure}[h!]
    \centering
    \begin{subfigure}
        {0.45\textwidth}
         \centering
         \includegraphics[width=\textwidth]{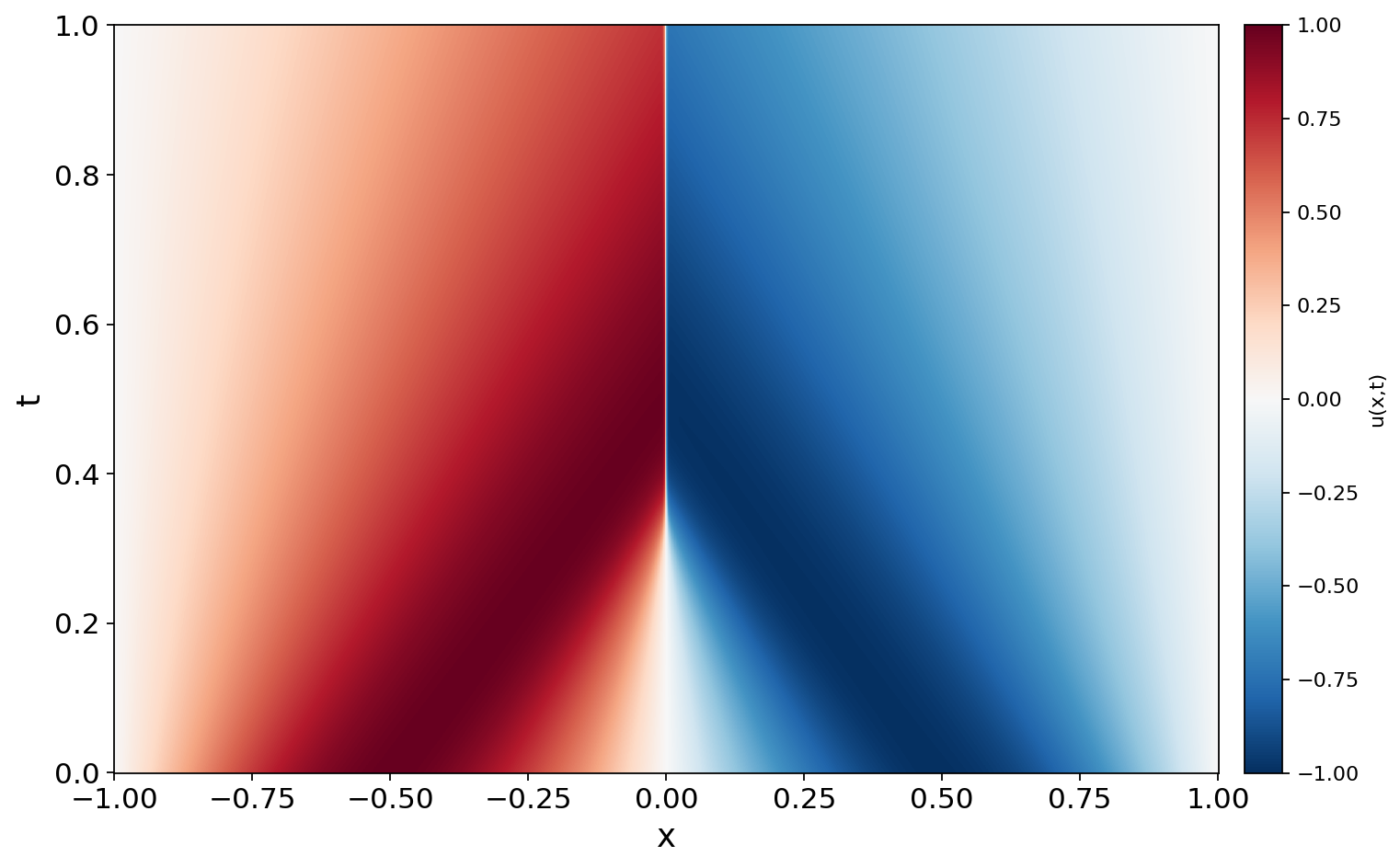}
         \caption{ }
         \label{fig:nu_001_gmm_pinn}
    \end{subfigure}
    \begin{subfigure}
        {0.45\textwidth}
         \centering
         \includegraphics[width=\textwidth]{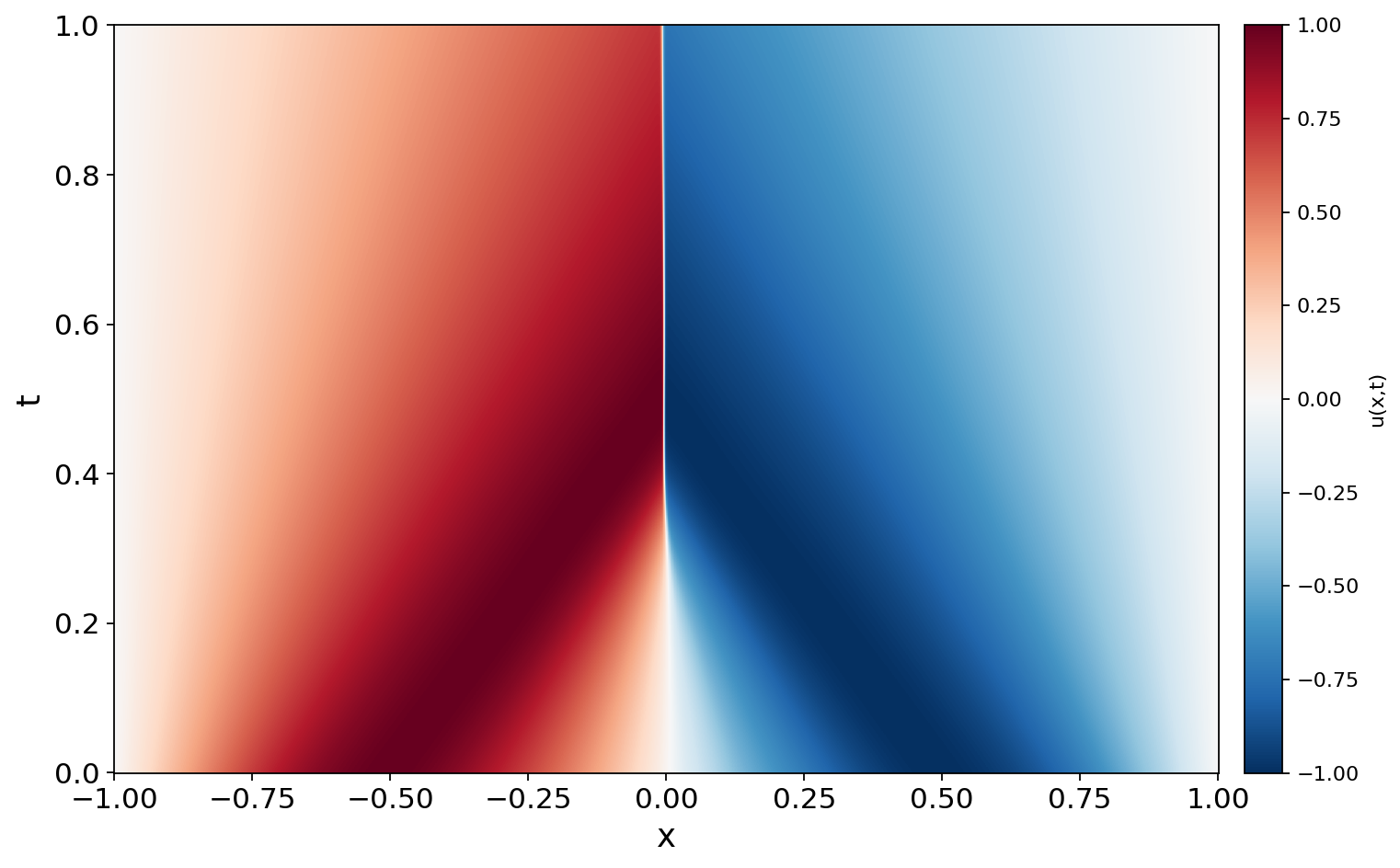}
         \caption{ }
         \label{fig:nu_001_pinn}
    \end{subfigure}
    \begin{subfigure}
        {0.5\textwidth}
         \centering
         \includegraphics[width=\textwidth]{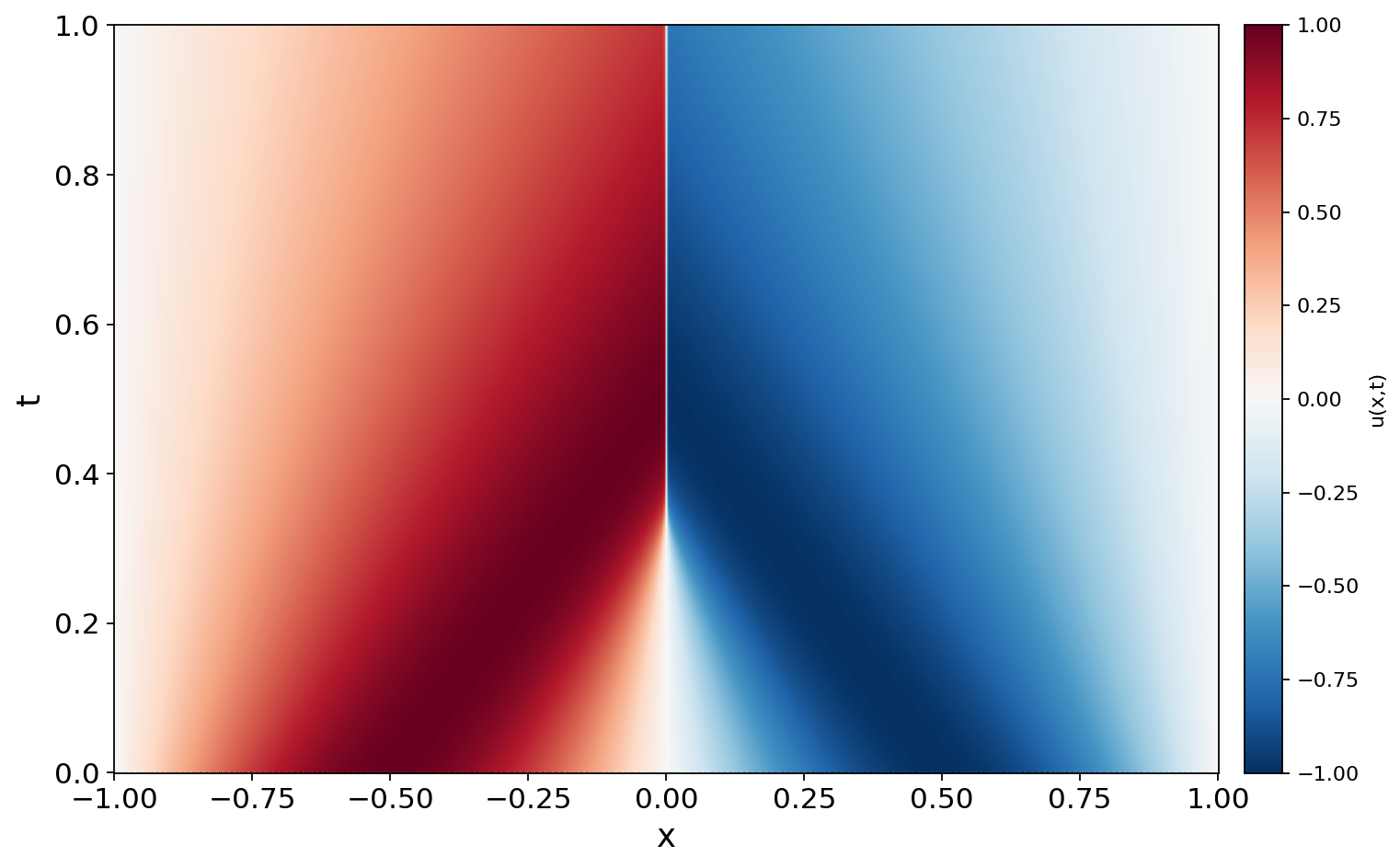}
         \caption{ }
         \label{fig:nu_001_ref}
    \end{subfigure}
    \caption{Solutions obtained using G-PINNs: Gaussian-based spatially weighted formulation \ref{fig:nu_001_gmm_pinn}, standard PINN \ref{fig:nu_001_pinn} and reference solution obtained with Cole-Hopf transformation \ref{fig:nu_001_ref}.}
    \label{fig: solution_nu_001}
\end{figure}

\begin{figure}
    \centering
    \includegraphics[width=0.8\linewidth]{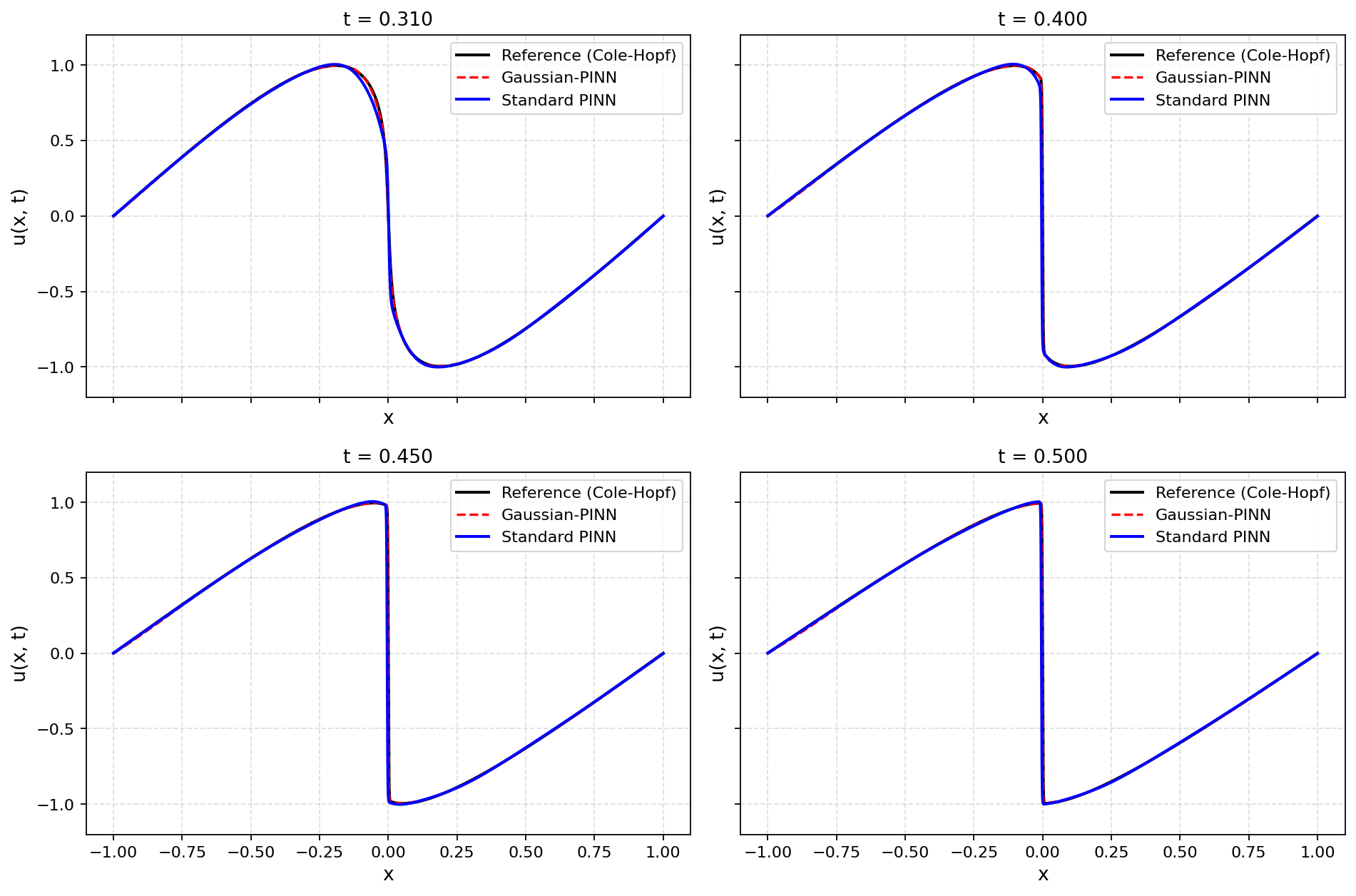}
    \caption{Solutions profiles at different time instants}
    \label{fig:nu_001 slices}
\end{figure}

\begin{figure}
    \centering
    \begin{subfigure}{0.49\textwidth}
    \includegraphics[width=\linewidth]{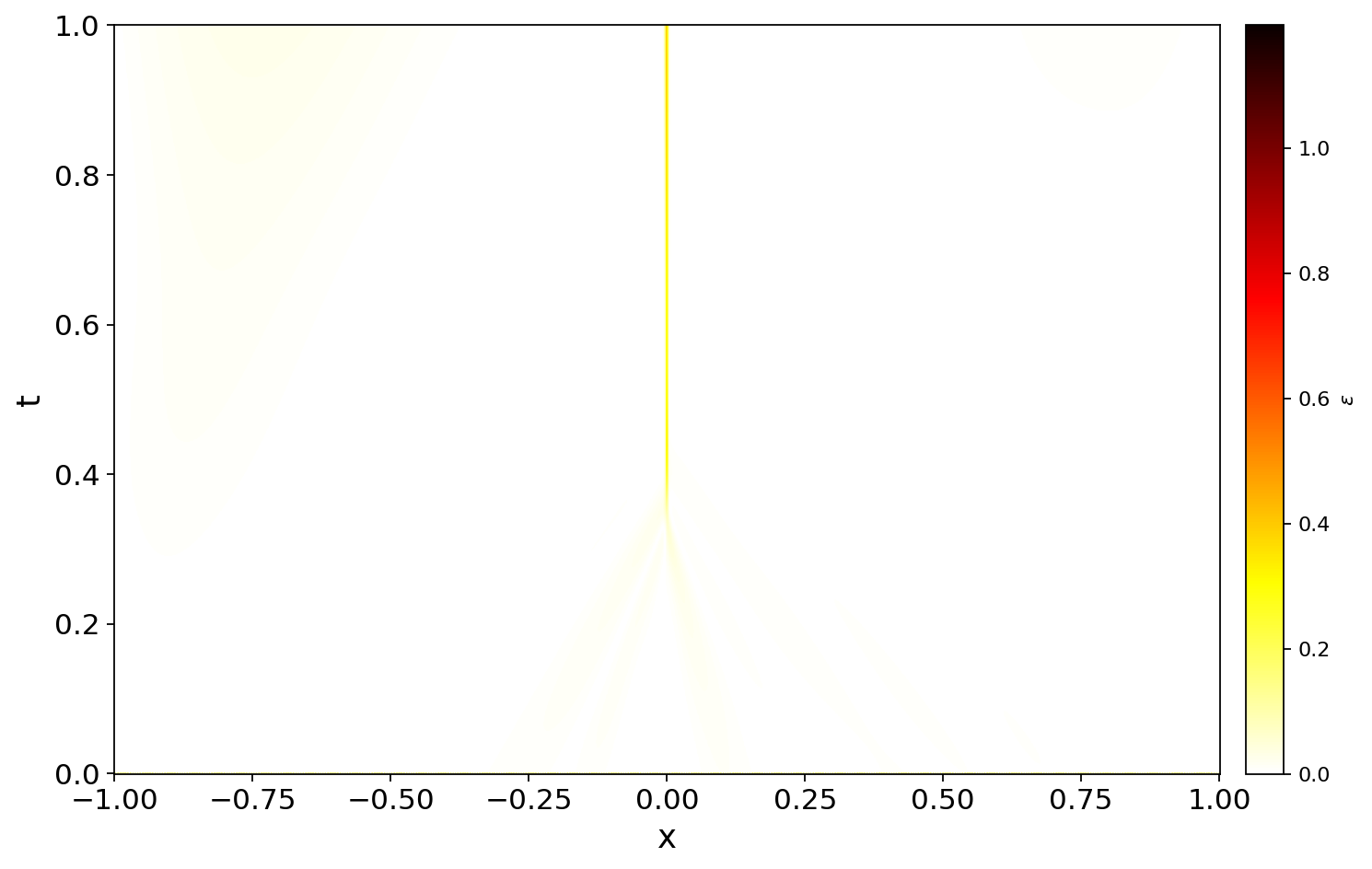}
    \caption{ }
    \label{fig: nu_001_error_gmm}    
    \end{subfigure}
        \begin{subfigure}{0.49\textwidth}
    \includegraphics[width=\linewidth]{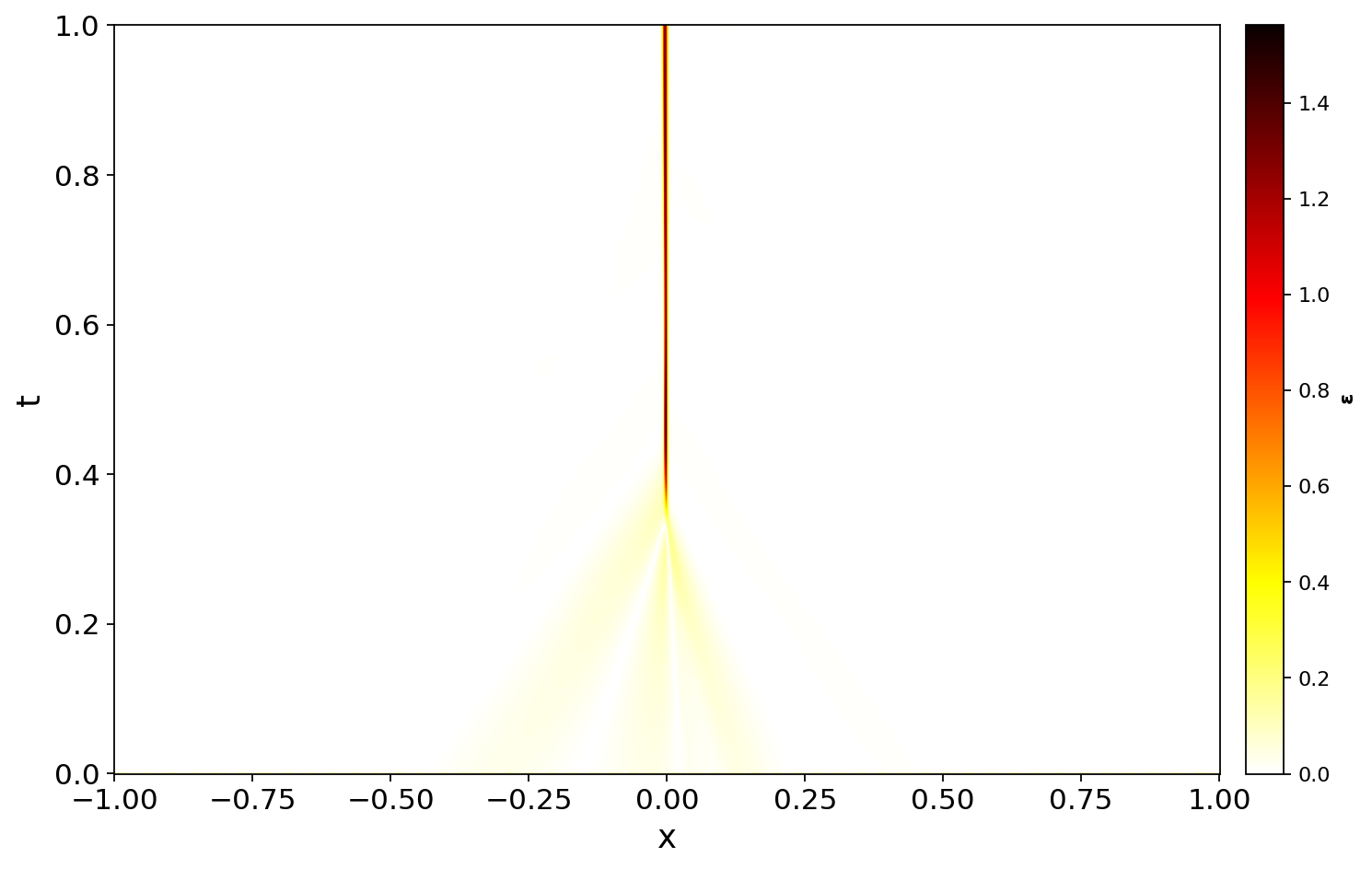}
    \caption{ }
    \label{fig: nu_001_error_pinn}    
    \end{subfigure}
    \caption{Absolute error maps of the G-PINNs: Gaussian-based spatially weighted loss formulation method \ref{fig: nu_001_error_gmm} and the standard PINN \ref{fig: nu_001_error_pinn}.}
    \label{fig: n_001_error}
\end{figure}

This section serves as complementary section of section \ref{burgers: 0.0005-static} . We consider a low viscosity Burgers equation ($\nu=0.001$) with a static shock wave:
$$
\begin{cases}
    u_t + uu_x = 0.001 u_{xx},  \quad x\in [-1,1], \quad t\in [0,1]\\
    u(x,0) = -\sin(\pi x) \\
    u(-1,t) = u(1,t) = 0.
\end{cases}
$$
For this value of $\nu$, a static vertical shock wave is known to form at $x=0$. Figure \ref{fig: solution_nu_001} illustrates the solutions obtained using the G-PINN methodology, the standard PINN, and the Cole-Hopf solution.
Figure \ref{fig:nu_001 slices} presents the solutions obtained at different time instants $t \in[0.31, 0.4, 0.45, 0.5]$. A comparison of temporal slices reveals that the proposed model maintains a sharp discontinuity, aligning almost perfectly with the reference solution at the different time instants, particularly at the $t=0.31$ time instant, which is theoretically close to the breaking time. Although the standard solution of PINN deviates from the solution in shock viscosity at $t=0.31$, these discrepancies persist at later times. 
Figure \ref{fig: n_001_error} presents the absolute error heatmaps obtained by computing:
$$
\epsilon = |u^*(x,t) - u^{\theta} (x,t)|
$$
where $u^*$ is the Cole-Hopf solution and $u^{\theta}$ is the solution obtained by standard PINN or G-PINN formulation. The error heat maps reveal the superior results obtained with the G-PINN formulation. Figure \ref{fig: nu_001_error_pinn} shows a maximum absolute error that reaches approximately $\epsilon \approx 1.5$ mostly concentrated along a sharp vertical line that coincides with the shock front at $x=0$. These errors progressively increased over time. Moreover, for $t<0.35$, a non-negligible error is observed in the neighbor of $x=0$. By contrast, figure \ref{fig: nu_001_error_gmm} reveals that the presented methodology could maintain the absolute error bounded by $\epsilon \approx 1$, with much lower error levels around the shock wave at $x=0$, these errors are much better balanced over time.  
Table \ref{table: nu_001} highlights the L2 relative error computed by:
$$
L2 = \frac{||u^* - u^{\theta}||}{||u^*||} 
$$
and the Mean Absolute Error (MAE) defined by:
$$
\text{MAE} = \frac{1}{N}\sum_{i=1}^{N} |u^*(x^i,t^i)-u^{\theta}(x^i,t^i)| 
$$
with: $N$ the total number of grid points. 
\begin{table}[h!]
    \centering
   \begin{tabular}{c|c|c}
   \hline
    Case & L2  & MAE \\
    \hline
    standard PINN & $9.14 \times 10^{-2}$ & $7.69\times 10^{-3}$ \\
    G-PINN & $3.71 \times 10^{-2}$ & $4.33\times 10^{-3}$ \\
    \hline
\end{tabular}
   \caption{Viscous Burgers' Eq. with $\nu = 0.001$: Error metrics}
    \label{table: nu_001}
\end{table}
The values obtained validate the superiority of the G-PINN methodology. The proposed model could achieve a relative L2 error of approximately $3.7\%$ compared to the $9.1\%$ error for standard PINNs. Similarly, the MAE values yielded a value of $4.33\times 10^{-3}$ for the presented methodology compared to $7.69\times 10^{-3}$ for the standard PINN. 
\section{Detected shock waves}
\begin{figure}[h!]
    \centering
    \begin{subfigure}{0.49\textwidth}
    \includegraphics[width=\linewidth]{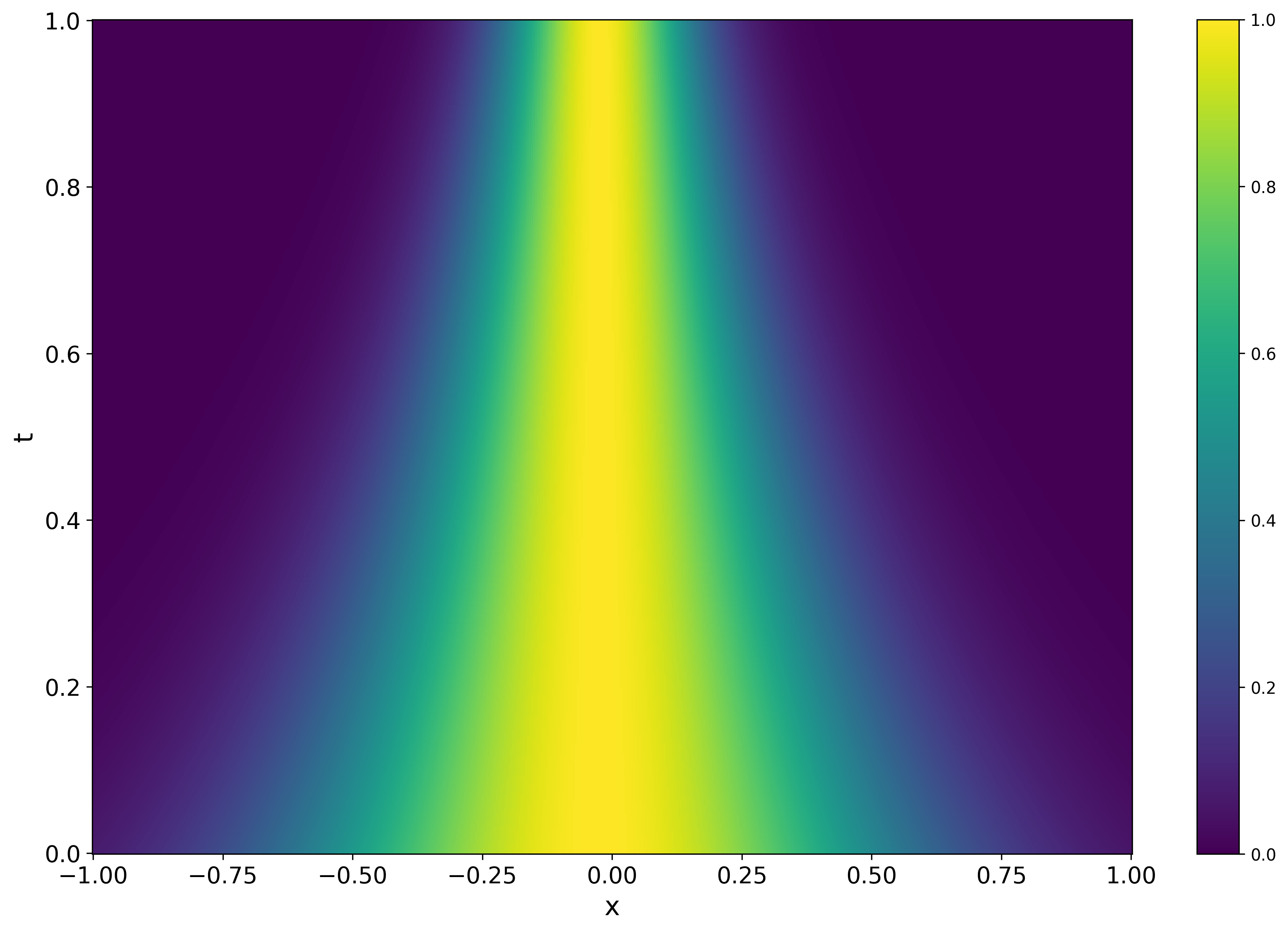}
    \caption{ }
    \label{fig: nu_0005_g_dist}  
    \end{subfigure}
    \hfill
        \begin{subfigure}{0.49\textwidth}
    \includegraphics[width=\linewidth]{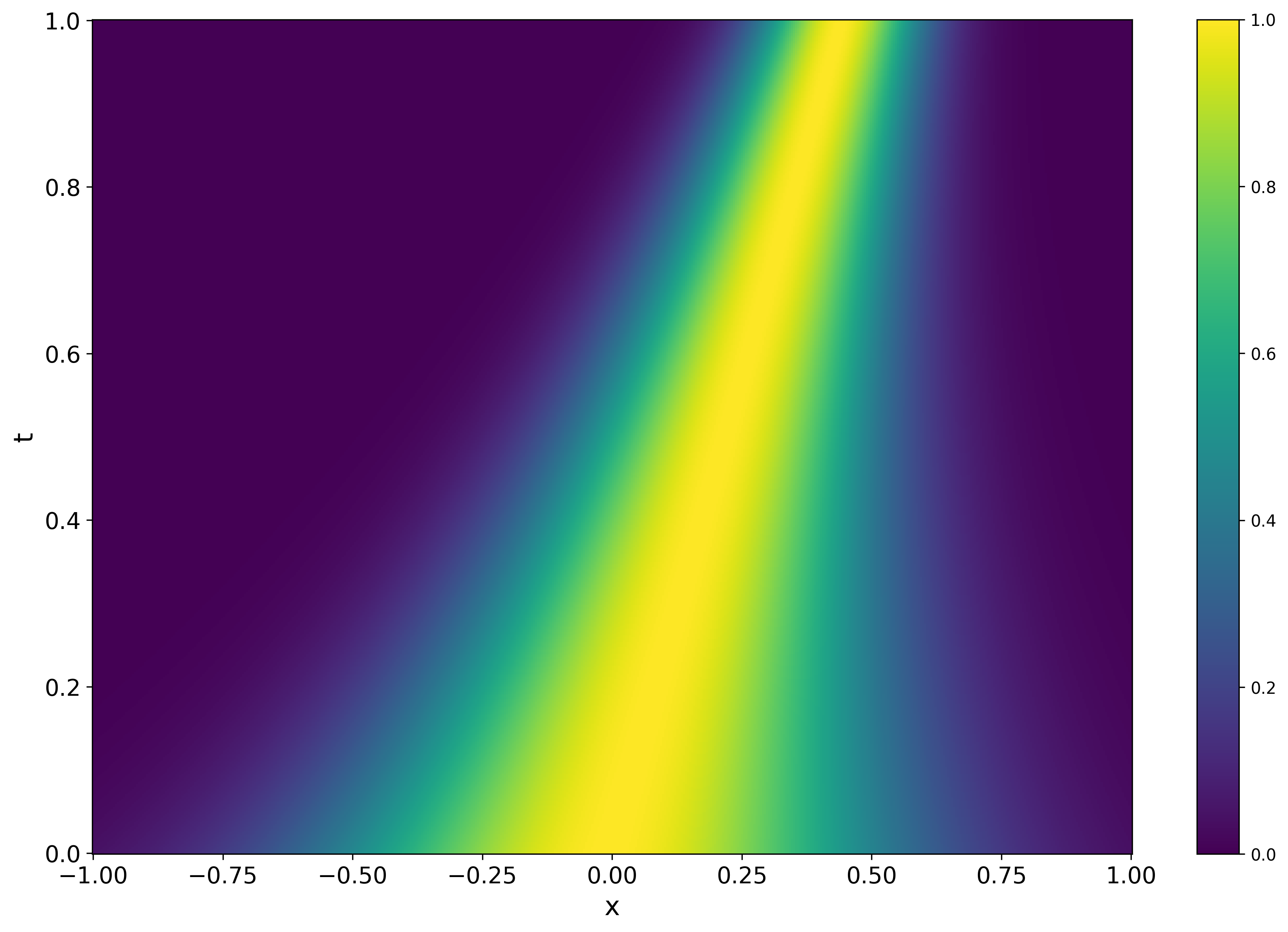}
    \caption{ }
    \label{fig: nu_0005_moving_g_dist}    
    \end{subfigure}
    \caption{Gaussian distributions obtained for the static shock wave and the moving shock wave with $\nu=0.0005$.}
    \label{fig: g_dist}
\end{figure}

Figure \ref{fig: g_dist} depicts the Gaussian distributions obtained for the static and moving shock waves test cases with $\nu=0.0005$ see sections \ref{subsec: static _shock_0005} and \ref{subsec: moving _shock_0005}, The learned  distribution, shown in Figure \ref{fig: nu_0005_g_dist}, concentrates its highest probability density in a  vertical line centered at $x \approx 0$
across all $t \in [0,1]$. This is consistent with the static shock of this test case: since the initial condition $u(x,0)= -\sin(\pi x)$
is an odd function, the anti-symmetry of the Burgers equation guarantees a zero shock speed. The learned shock speed $m \approx 0.001$ is in excellent agreement with the theoretical value, demonstrating that the proposed method has correctly identified the region of the highest residual PDE. For the moving shock case $u(x,0)= -\sin(\pi x) + 0.5$ shown in figure \ref{fig: nu_0005_moving_g_dist}, the learned distribution tilts with a  slope $m \approx 0.45-0.49$ in 
the $(x,t)$ plane, automatically tracking the trajectory of the shock, without prior knowledge of the location of the shock. 

\section{Cole-Hopf solution of viscous Burgers equation} \label{appendix: cole-Hopf} 
The Cole-Hopf formulation \cite{Hopf1950ThePD} is used to solve the viscous Burgers equation given by:
\begin{equation} \label{eq: viscous_eq}
\frac{\partial u}{\partial t} + u \frac{\partial u}{\partial x} = \nu \frac{\partial^2 u}{\partial x^2},   
\end{equation}
where $u$ is the velocity scalar field and $\nu$ is the kinematic viscosity. The Cole-Hopf method relies on the variable change: 
\begin{equation} \label{eq: var_change}
u(x,t) = -2 \nu \frac{\partial  \ln \Phi(x,t)}{\partial x} = -2\nu\frac{\Phi_x}{\Phi}.    
\end{equation}
By substituting \ref{eq:deviation} into \ref{eq: viscous_eq}, we obtain the following linearized heat equation:
\begin{equation}\label{eq: heat}
    \frac{\partial \Phi}{\partial t} = \nu \frac{\partial^2 \Phi}{\partial x^2}.
\end{equation}

For an initial condition of \ref{eq: viscous_eq} $u(x,0)= u_0$, we have:
$$
\Phi(x,0) = e^{-\frac{1}{2\nu} \int_{-\infty}^0 u(\eta) d \eta  },
$$
subsequently, the general solution of the heat equation \ref{eq: heat} is given by: 
\begin{equation} \label{eq: solution_heat}
\Phi(x,t) = \frac{1}{\sqrt{4\pi\nu t}} \int_{-\infty}^{\infty} \Phi(\eta,0) \exp \left( -\frac{(x-\eta)^2}{4\nu t} \right) d\eta.   
\end{equation}

Substituting \ref{eq: solution_heat} into \ref{eq: var_change}, we obtain:
\begin{equation}
    u(x,t) = \frac{\int_{-\infty}^{\infty} \frac{x-\eta}{t} \exp \left( -\frac{G(\eta, x, t)}{2\nu} \right) d\eta}{\int_{-\infty}^{\infty} \exp \left( -\frac{G(\eta, x, t)}{2\nu} \right) d\eta},
\end{equation}
where: 
$$
G(\eta, x, t) = \int_{0}^{\eta} u_0(\zeta) \, d\zeta + \frac{(x-\eta)^2}{2t}.
$$
Although the Cole-Hopf transformation provides an analytical solution of the Burgers equation, its numerical evaluation for the purpose of this study relies on a truncated series expansion. All the error metrics reported in this work are computed in relation to this numerical solution.


\end{document}